\documentclass[aps,pre,twocolumn,groupedaddress,floatfix,longbibliography,superscriptaddress]{revtex4-1}
\usepackage{amsmath}
\usepackage{amssymb}
\usepackage{amsfonts}
\usepackage{ulem}
\usepackage{graphicx}
\usepackage{bbold}
\usepackage{bm}
\usepackage{bm}
\usepackage{cancel}
\usepackage{times,float}
\usepackage{graphicx}
\usepackage{float}
\usepackage[usenames,dvipsnames,svgnames]{xcolor}
\usepackage{hyperref}
\hypersetup{colorlinks=true, linkcolor=Blue, citecolor=PineGreen,urlcolor=cyan}
\usepackage{multirow}
\usepackage{textcomp}

\newcommand{\bea}{\begin{eqnarray}}
\newcommand{\eea}{\end{eqnarray}}
\newcommand{\nn}{\nonumber}

\begin{document}
%____________________________________
\title{Computing the gravitational acceleration within the World Pendulum Alliance: an application of the remote laboratory methodology implemented by UNAD }
%____________________________________
\author{Freddy Torres-Payoma}
\email{freddy.torres@unad.edu.co}
\address{Universidad Nacional Abierta y a Distancia - UNAD, Escuela de Ciencias Básicas, Tecnología e Ingeniería, Bogotá D.C,111511, Colombia}
%____________________________________
\author{Diana Herrera}
%\email{diana.herrera@unad.edu.co}
\address{Universidad Nacional Abierta y a Distancia - UNAD, Escuela de Ciencias Básicas, Tecnología e Ingeniería, Bogotá D.C,111511, Colombia}
%____________________________________
\author{Karla Triana}
%\email{karla.triana@unad.edu.co}
\address{Universidad Nacional Abierta y a Distancia - UNAD, Escuela de Ciencias Básicas, Tecnología e Ingeniería, Bogotá D.C,111511, Colombia}
%____________________________________
\author{Laura Neira-Quintero}
%\email{ldneiraq@unad.edu.co}
\address{Universidad Nacional Abierta y a Distancia - UNAD, Escuela de Ciencias Básicas, Tecnología e Ingeniería, Bogotá D.C,111511, Colombia}
%____________________________________
\author{Jorge David Casta\~no-Yepes}
\email{jcastano@uc.cl}
\address{Universidad Nacional Abierta y a Distancia - UNAD, Escuela de Ciencias Básicas, Tecnología e Ingeniería, Bogotá D.C,111511, Colombia}
\affiliation{Instituto de Física, Pontificia Universidad Católica de Chile, Vicuña Mackenna 4860, Santiago, Chile.}
%____________________________________ 

%%%%%%%%%%%%%%%%%%%%%%%
%
\begin{abstract}
In this paper, we calculate the gravitational acceleration by using simple pendulums within the designated World Pendulum Alliance: a network constituted by fourteen institutions in eight countries, each provided by a pendulum that can be accessed remotely via the Internet. As a pedagogical option for remote laboratory experiences, we show how to access them and measure the pendulum's period for $N$ oscillations. We discuss the role of the repetitions and samples to minimize the experimental uncertainty, and by using the averaged value of the period, we compute the gravitational acceleration locally. Also, we present the corrections due to the moment of inertia of the system. Finally, by taking into account the geographical location of each pendulum, we report gravity's dependence on latitude, longitude, and altitude by discussing the implications on the Earth's shape. We hope this tool will be helpful for introductory and university physics courses due to its remote access features and its relation to fundamental concepts.
\end{abstract}
\pacs{xxx xxx xx xx}
\maketitle

%%%%%%%%%%%%%%%%%%%%%%%%%%%%%%%%%%%%%%%%%%%
\section{Introduction}
%%%%%%%%%%%%%%%%%%%%%%%%%%%%%%%%%%%%%%%%%%%
The global health situation for COVID19 has changed the public and private education paradigms. The restrictive quarantines and the lack of strategies for remote or virtual teaching methodologies slowed down access to knowledge in primary and higher education, being the basic and professional laboratory practices the most affected activities. The latter highlighted the necessity to have and develop new tools to bring the whole experimental experiences to students, not only while the pandemic remains but also to generate alternatives that can be implemented for those who cannot study in a presential system. By the hand of this, nowadays the Remote-Controlled Laboratories (RCL) and interactive/remote teaching are a necessity for the new generations. The young students are more familiar with strategies of self-learning~\cite{PhysRevSTPER.11.010104,PhysRevSTPER.3.010101}, which {include}  interactive screens~\cite{Kirstein_2007}, interactive videos~\cite{Wagner_2006}, and experiments accessible via Internet~\cite{Gr_ber_2007}. 

In Colombia, the Universidad Nacional Abierta y a Distancia (UNAD) is one of the pioneer centers {in offering} higher education with virtual and distance modalities, with more than 135.000 students in its different education systems. One of its primary purposes is {\it contributing to education for all through the open and distance modality}, overcoming the barriers that arise from the country's socioeconomic differences~\cite{afanador2021educacion}. Since most of the students of UNAD belong to the less privileged social classes, some of these limits are related to access to education in remote areas hardly connected by transport services to main cities with better equipment for STEM (science, technology, engineering, and mathematics) education. As a virtual and distance education university, UNAD faces a challenge of permanent innovation to incorporate new technologies that allow the curriculum to foster autonomous, meaningful, and collaborative learning. The objective is to educate competent students in any geographic region to promote sustainable economic, social and human development.

In the spirit of the above, UNAD has implemented a series of online experiments in physics within the World Pendulum Alliance (WPA@elab) network, which connects fourteen institutions in eight countries with a net of pendulums that are remotely accessible. These RCL {allows us to} measure in real-time the pendulum dynamics and calculate the variations of the gravitational acceleration as a function of the geographic coordinates. {This} RCL network was initially developed by the Instituto Superior Técnico (IST) by introducing a Massive Open Online Course (MOOC) in 2016 as an alpha-version accessible to its students~\cite{torres2019collaborative}. In 2019, UNAD joined the project by evaluating and correcting a beta version focusing efforts in an RCL devoted uniquely to oscillatory phenomena inside the MOOC.

From a basic science point of view, studying the changes of the Earth's gravitational field (EGF) is an active area of research. Several collaborations have worked on mapping the variations of gravity around the globe. For instance, the {\it Gravity field and steady-state Ocean Circulation Explorer} (GOCE) with satellite gradiometry, {have} observed the EGF from space with unprecedented resolution~\cite{Drinkwater2003,rebhan2000gravity}. It has shown the gravitational field profile due to the variations of the shape of Earth and modeling its crust at regional and global scales~\cite{Sampietro2016}. Also, the {\it Gravity Recovery and Climate Experiment}~\cite{GRACE3} measured the EGF and inferred important Earth's features like regional surface-mass anomalies~\cite{GRACE1}, models for groundwater distribution~\cite{GRACE2}, and global ocean mass variations~\cite{GRACE4}. Moreover, the Earth is not a static system given the tectonic plates evolution~\cite{turcotte2002geodynamics} (which is an open problem~\cite{chen2019southward,boonma2019lithospheric,MORENO2021103604}), making the EGF sensitive to such a motion~\cite{dumberry2004variations}.

The above discussion, while qualitatively understandable, is not quantitatively accessible to the students of introductory and university physics, given the level of specialization of the methods, mathematical models, and computational details. Nevertheless, one may ask: {\it is it possible, from a pedagogical perspective, to perform simple measurements to infer EGF variations?} Then, this paper focuses on calculating the gravitational acceleration by using the WPA as an education alternative that uses RCL to test the gravity variations by comparing the data taken from the different pendulums of the alliance.
 This idea was proposed by Gröber {\it et al.}, who measured the gravitational acceleration in Speyer,
Germany~\cite{Gr_ber_2007}. In contrast, {within} the WPA, we report a data from eight countries and several altitudes. 

The paper is organized as follows: Sec.~\ref{Sec:The_WPA} gives a general panorama of the WPA and the relevant information {on} the pendulums. Then, Sec.~\ref{Sec:Fundamental_Concepts} presents a short review of the main fundamental concepts regarding the mathematical and physical pendulums, as well as a discussion about the geographical variations of gravitational acceleration. Next, in Sec.~\ref{sec:data_acquisition} we explain the data acquisition, and in Sec.~\ref{sec:Results_and_discussion} we discuss how to obtain the gravity from the measured period and its dependence on latitude, longitude, and altitude. Finally, we summarize and conclude in Sec.~\ref{sec:Summary and Conclusions}.

%%%%%%%%%%%%%%%%%%%%%%%%%%%%%%%%%%%%%%%%%%%
\section{The WPA}\label{Sec:The_WPA}
%%%%%%%%%%%%%%%%%%%%%%%%%%%%%%%%%%%%%%%%%%%
The World Pendulum Alliance (WPA) is a project co-financed by the Erasmus+ Program of the European Union. It articulates higher education institutions from Europe (Portugal, Spain, France, Czech Republic) and Latin America (Colombia, Panamá, Chile, Brazil) intending to improve mathematics and science education access and quality. At the end of 2021, participating universities have installed in their facilities 11 primary pendulums available for anyone in the world with internet access. In addition, they are currently working on deploying 120 secondary pendulums in different places spread across the alliance's member countries. The network {\it enables teachers and students to collect experimental data in real-time at a planetary scale and measure one (or more) of the earth's physical characteristics on their own: the variation of gravity with latitude}~\cite{escobar2019pendulum}. Part of the Project's goals is to promote the local use of the network at national and international levels for developing resources to foster research and didactics for experimental sciences and engineering. 

One of the main objectives of the WPA is the internationalization of higher education. Especially the Project reached the internationalization of the curriculum on different fronts:
\begin{itemize}
\item Intercultural exchange of teachers during project implementation, increasing proficiency in the language (face to face and virtual meetings).
    
    \item Collaborative online international teaching by creating free access educational content to students in massive open online {courses} (MOOCs) and GRAASPs, a digital educational platform.
    
    \item Collaborative online international learning of students using the remote pendulum laboratory and access to educational platforms (MOOC and GRAASPS).
\end{itemize}

Moreover, the Project impacts other educational levels because the universities belonging to the WPA have secondary education institutions as allies. Furthermore, it brings digital education with broad geographical coverage and equality in developing countries. In this spirit, the students can improve their autonomous learning skills through remote experimentation, facilitating physics teaching at different levels of education, including secondary school.

%--------------------------------------------------
\begin{figure}[H]
    \centering
    \includegraphics[scale=0.45]{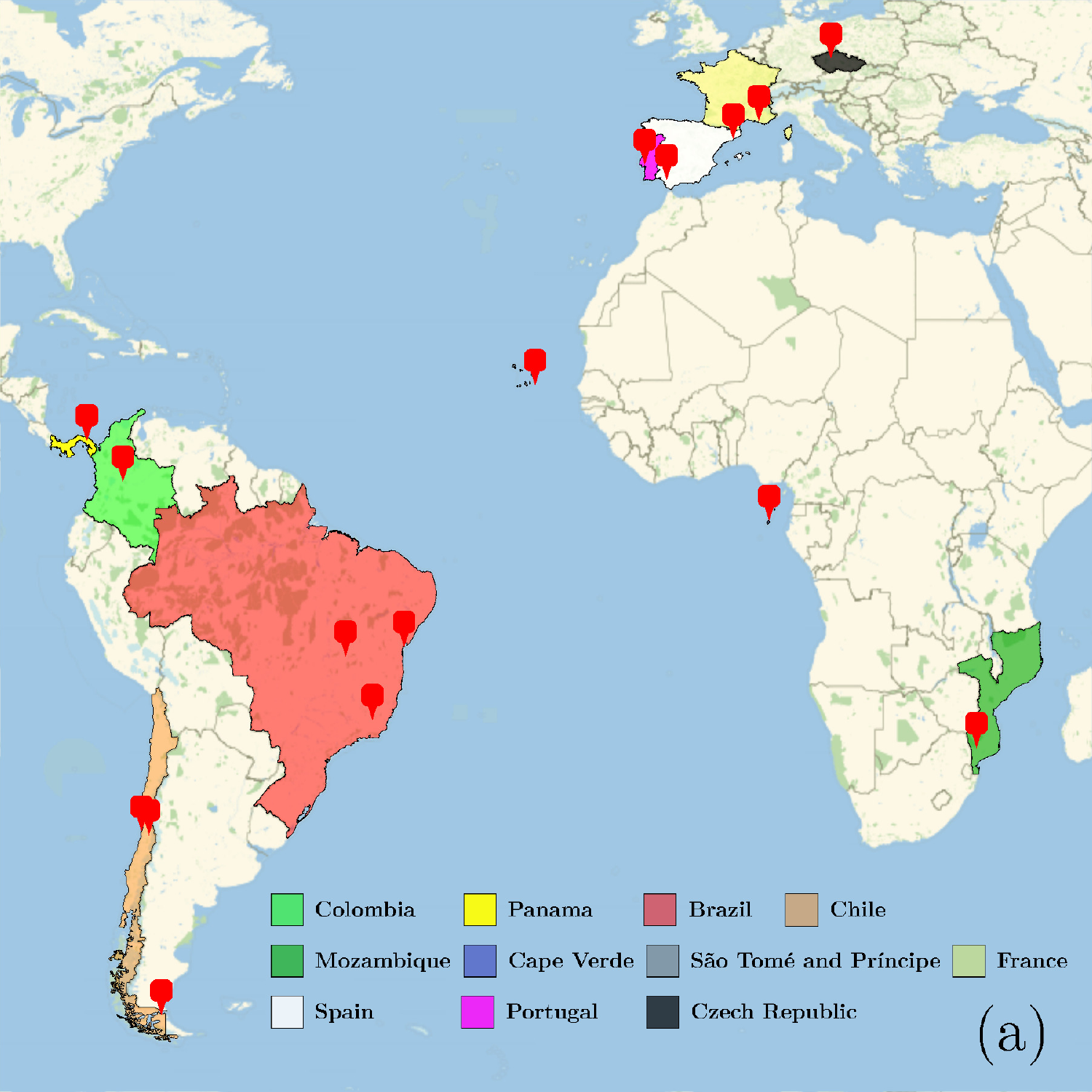}\\
    \vspace{0.5cm}
    \hspace{0.865cm}\includegraphics[scale=0.45]{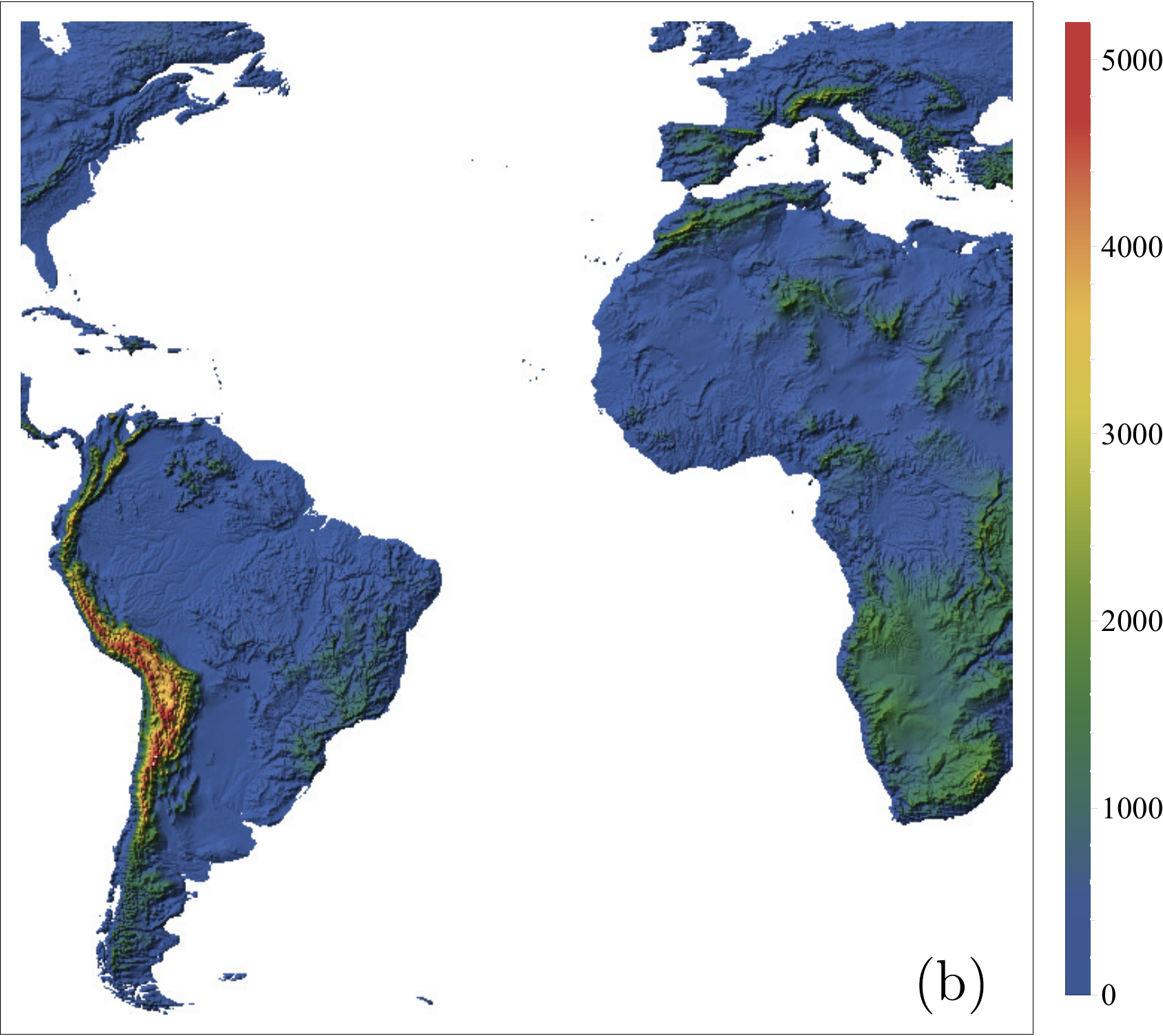}
    \caption{The WPA: (a) Geographical locations of the pendulums. In South-America: Colombia, Panamá, Brazil, Chile. In África: Mozambique, São Tomé and Príncipe, Cape Verde. In Europe: France, Spain, Portugal, Czech Republic. (b) Detail of the relief. The color code and bar indicate the altitude in meters.}
    \label{fig:locations}
\end{figure}
%--------------------------------------------------

Figure~\ref{fig:locations}(a) shows the locations of the 11 primary pendulums around the world. As complement, in Fig.~\ref{fig:locations}(b) the relief of each location is displayed. The information of these figures suggests that within the WPA, the students can access remote experiments with variations in latitude and altitude, which constitutes a novel way to obtain experimental data in real-time. The exact {primary} pendulum's locations, and their physical parameters are depicted in Table~\ref{tab:ExperimentalParameters}.
\newpage
\onecolumngrid
\begin{table*}
    \centering
    \begin{tabular}{|c|c|c|c|c|c|c}
    \hline
{\bf Location} &	{\bf Latitude}	& {\bf Longitude} & {\bf Altitude [m]} & {\bf String length [mm]} &	{\bf Sphere diameter [mm]}\\
\hline
CCV Algarve/Faro & 37º00'N	& 7º56'O & 10 &	2677 $\pm$ 0,5 (at 23ºC) & 80,5 $\pm$ 1,0\\
\hline
UESC/Ilhéus & 14º47'S & 39º10'O & 220 & 2832.0 $\pm$ 0.5 (at 23ºC) & 81,0 $\pm$ 1,0\\
\hline
Lisbon & 38º41'N & 9º12'O & 20 & 2677 $\pm$ 0.5 (at 19ºC) & 80,5 $\pm$ 1,0\\
\hline
Maputo & 25º56'S & 32º36'E & 80 & 2609.8 $\pm$ 0.5 (at 27ºC) & 80,5 $\pm$ 1,0\\
\hline
Santo Tomé & 0º21'N & 6º43'E & 50 & 2756.5 $\pm$ 0.5 (at 29ºC) & 81,8 $\pm$ 0,5\\
\hline
Prague - CTU & 50º5.5'N & 14º25.0'E & 150 & 2803 $\pm$ 0,5 (at 25ºC) & 80,1 $\pm$ 0,5\\
\hline
Barcelona-UPC & 41º24.6'N & 2º13.1'E & 55 & 2824 $\pm$ 1 & 81,8 $\pm$ 0,1\\
\hline
Río de Janeiro - PUC & 22º54.1'S & 43º12'O & 50 & 2826,0 $\pm$ 0,5 & 81,6 $\pm$ 0,1\\
\hline
Praia - UniCV & 14°56'N & 23°31'O & 40 & 2832,0 $\pm$ 0,5 & 81,6 $\pm$ 0,1\\
\hline
Bogotá - UniAndes & 4°36'N & 74°3'O & 2500 & 2824 $\pm$ 0,5 & 82,0 $\pm$ 0,1\\
\hline
Bogotá - UNAD & 4°35'N & 74°5'O & 2578 & 2835 $\pm$ 0,5 & 82,0 $\pm$ 0,1\\
\hline
Panamá City - UTP & 9°1.3'N & 79°31.9'O & 82 & 2825 $\pm$ 0,5 (at 28ºC) & 81,9 $\pm$ 0,1\\
\hline
Santiago - UChile & 33°27.5'S & 70°39.8'O & 552 & 2825 $\pm$ 0,5 (at 27ºC) & 81,9 $\pm$ 0,1\\
\hline
Valparaíso - UTFSM & 33°1'S & 71°37'O & 30 & 2827.5 $\pm$ 0.5 (at 28ºC) & 81,8 $\pm$ 0,1\\
\hline
Panamá City - USMA & 9°1'N & 79°37'O & 130 & 2800.0 $\pm$ 0.5 (at 35ºC) & 81,8 $\pm$ 0,1\\
\hline
Brasília - UnB & 15° 46'S & 47° 52'O & 1034 & 2826.8 $\pm$ 0.5 (at 26ºC) & 81,4 $\pm$ 0,1\\
\hline
Marseille - ECM & 43°20.6'N & 5°26.2'E & 162 & 2828.0 $\pm$ 0.5 (at 22ºC) & 82,0 $\pm$ 0,1\\
\hline
Punta Arenas - UMag & 53°8'S & 70°52'O & 40 & 2823 $\pm$ 0.5 @16.4ºC) & 81,7 $\pm$ 0,1\\
\hline
    \end{tabular}
    \caption{Locations and parameters of the WPA's primary pendulums~\cite{WinNT}.}
    \label{tab:ExperimentalParameters}
\end{table*}
\twocolumngrid

\begin{figure}[h!]
    \centering
    \includegraphics[scale=0.49]{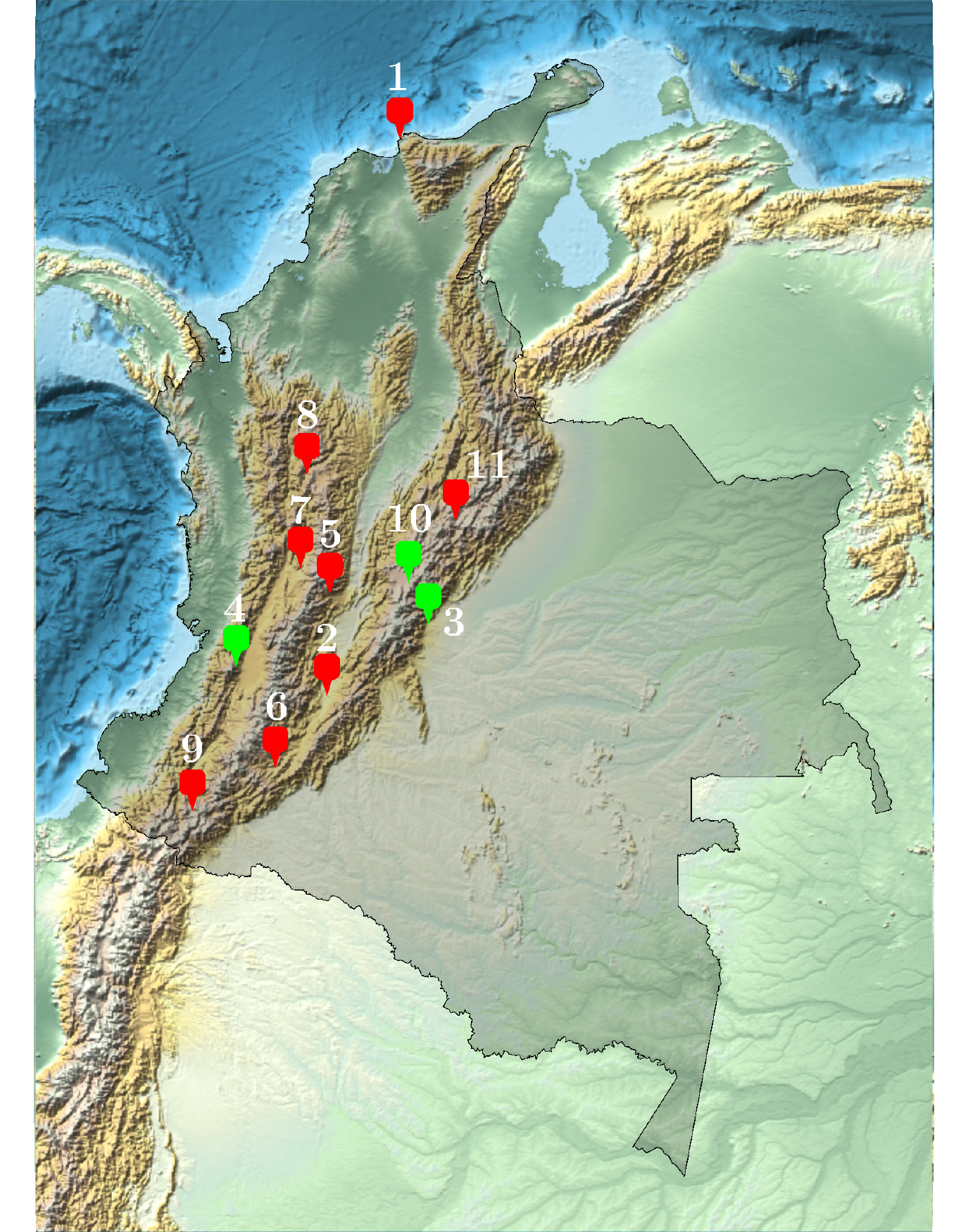}
    \caption{(a) Secondary pendulum's locations in Colombia (WPA@UNAD network 2022): 1. Santa Marta (Magdalena), 2. Neiva (Huila), 3. Acacías (Meta), 4. Cali (Valle del Cauca), 5. Ibagué (Tolima), 6. Pitalito (Huila), 7. Dosquebradas (Risaralda), 8.Medellín (Antioquia), 9. Pasto (Nariño), 10. Bogotá D.C., 11. Tunja (Boyacá). The relief is shown in order to appreciate the altitude changes. The green markers are the installed pendulums, whereas the red ones are the locations for future installations.}
    \label{Fig:pendulum_map_1}
\end{figure}

\begin{figure}[h!]
    \centering
    \includegraphics[scale=0.58]{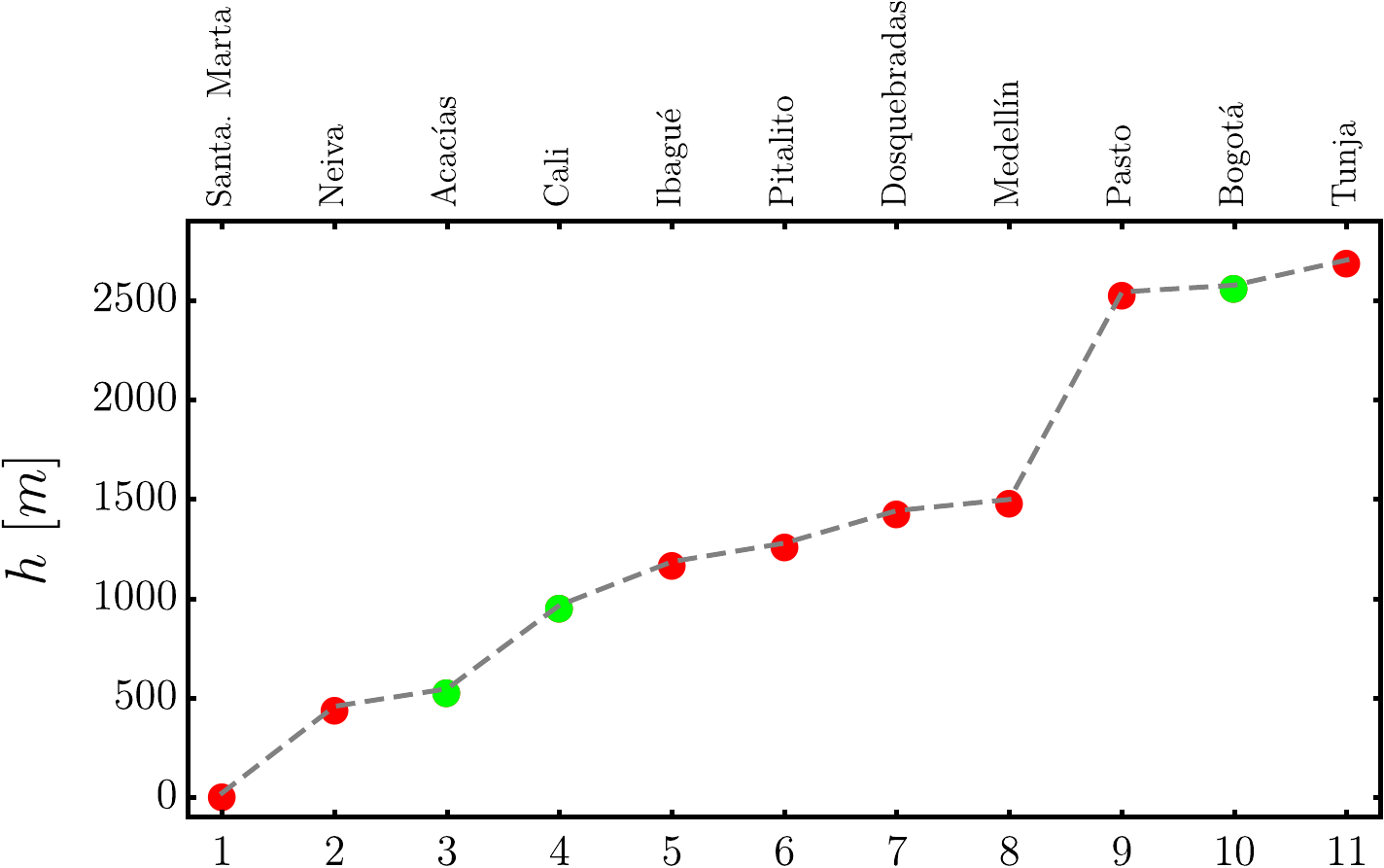}
    \caption{Altitude $h$ in meters for each secondary pendulum in Colombia. The green markers are the installed pendulums, whereas the red ones are the locations for future installations.}
    \label{Fig:pendulum_map_2}
\end{figure}

The first phase of the project underlies the installation of the primary pendulum network, which is the focus of this article. The next phase of the WPA project at UNAD started at the end of 2021 with installing the first secondary wall pendulum at the University's main campus in Bogotá. The project consists of installing 11 secondary pendulums under similar installation and calibration criteria of wall pendulums whose dimension does not exceed 2 m in length. In addition, some of the UNAD's campuses nationwide were selected for the installation. As a result, the local collaboration has installed three secondary pendulums (Bogotá, Acacias, and Cali). The next pendulums will be installed at the UNAD sites in Santa Marta, Ibague, Dosquebradas, Medellín, Pasto, Tunja, Pitalito, and Neiva, covering a large part of the Colombian territory as is shown in Fig.~\ref{Fig:pendulum_map_1}. After installation, the research aims to evaluate possible gravitational effects due to elevation. Therefore, pendulums will be located at different heights, from 23 to 2070 meters above sea level. In Fig.~\ref{Fig:pendulum_map_2}, the altitude of the proposed locations is depicted.

% \begin{figure}[H]
%     \centering
%     \includegraphics[scale=0.5]{secundary_pendulum-eps-converted-to.pdf}
%     \caption{Constellation WPA@UNAD. The graph represents elevations at sea level. The minimum height in Santa Marta is 23 m and the maximum in Tunja is 1770 m.  }
%     \label{fig:pendulum_height}
% \end{figure}

%%%%%%%%%%%%%%%%%%%%%%%%%%%%%%%%%%%%%%%%%%%
\section{Fundamental concepts}\label{Sec:Fundamental_Concepts}
%%%%%%%%%%%%%%%%%%%%%%%%%%%%%%%%%%%%%%%%%%%
This section discusses the fundamental mathematical expressions to analyze the experimental data obtained with the WPA. It does not pretend to be a complete review of the subjects, and the interested reader can consult the provided references. Also, we briefly discuss the importance of the Earth's shape in order to present our results in terms of the geographical position of each pendulum in the network.
%%%%%%%%%%%%%%%%%%%%%%%%%%%%%%%%%%%%%%%%%%%
\subsection{Basics of the mathematical and physical pendulum}
%%%%%%%%%%%%%%%%%%%%%%%%%%%%%%%%%%%%%%%%%%%Latitude-dependent gravitational acceleration

One of the simplest models to analyze in Newtonian and Lagrangian mechanics is the pendulum. Its study covers courses of introductory and university physics, so that one passes from the description of the so-called {\it mathematical pendulum} (a geometrical point with mass $m$ hanging from a massless wire) to the analysis of the {\it physical pendulum}, where the dimensions and moments of inertia of the bob and wire are taken into account. In both circumstances (by ignoring the dissipation of energy) the equation of motion for the bob/oscillating body can be found, which takes the form~\cite{landau1973classical,alonso1967fundamental,reese1998university}:
\bea
\frac{d^2\theta}{dt^2}=-\omega^2\sin\theta,
\label{MASeq}
\eea
where $\theta(t)$ is the {body's angular position}, and $\omega$ is a constant that depends on the value of the local gravitational acceleration and the physical characteristics of the pendulum (such as length and moment of inertia). The solution of Eq.~(\ref{MASeq}) provides the expression of the pendulum's period (the time to perform a complete oscillation), which can be achieved from several approximations. This work analyzes the following forms of the period $T$:
\begin{itemize}
    \item The simple mathematical pendulum with small initial amplitude, i.e., $\theta_0\ll1$ rad:
    \bea
    T\approx2\pi\sqrt{\frac{l}{g}}.
    \label{T0}
    \eea
    
    % \item Corrections for arbitrary initial amplitude in a mathematical pendulum:
    % %
    % \bea
    % T=2\pi\sqrt{\frac{l}{g}}\sum_{n=0}^\infty\left[\frac{(2n)!}{2^{2n}(n!)^2}\right]^2\sin^{2n}\left(\frac{\theta_0}{2}\right)
    % \eea
    
    \item The physical pendulum composed of a rigid uniform sphere of radius $R$ and a massless wire in the small initial amplitude approximation:
    \bea
    T\approx2\pi\sqrt{\frac{l}{g}\left(1+\frac{2R^2}{5l^2}\right)},
    \label{T1}
    \eea
\end{itemize}
where $l$ {represents} the wire's length and $g$ is the local gravitational acceleration.

For the WPA, the validity of small oscillations approximation can be analyzed from the data of Table~\ref{tab:ExperimentalParameters}. There, the average wire's length is $l\approx 280$cm, and for a horizontal sphere displacement $x_0$, the initial angle $\theta_0$ is given by
\bea
\theta_0=\arcsin(x_0/l).
\label{theta0vsX0}
\eea

\begin{figure}[H]
    \centering
    \includegraphics[scale=0.6]{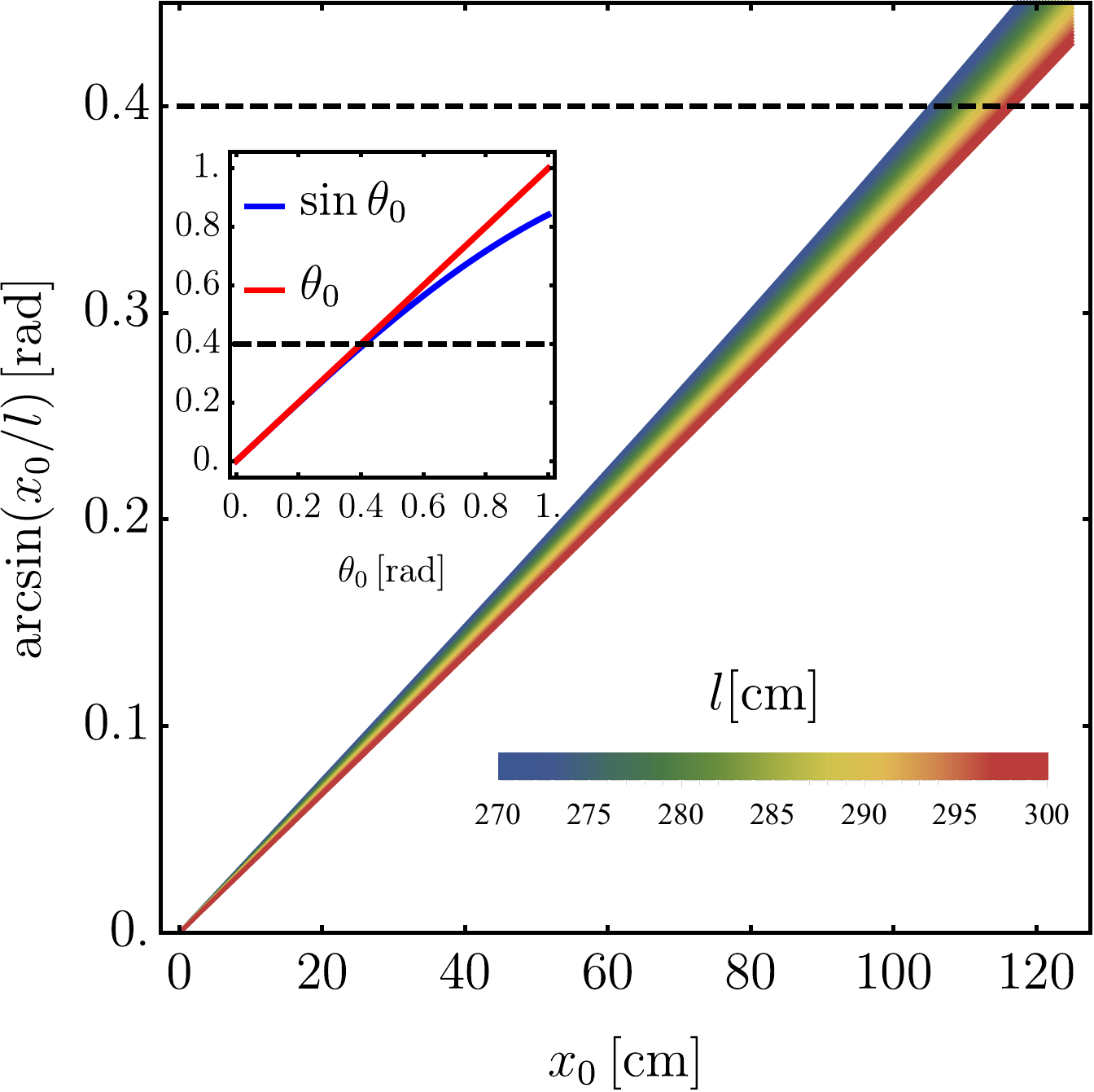}
    \caption{{The domain} of initial displacement $x_0$ where the approximation of small oscillations is valid (below the dashed line $\theta_0\simeq 0.4$). The color bar indicates a set of {string lengths} $l$ close to the parameters of Table~\ref{tab:ExperimentalParameters}. The inset shows the values of $\theta_0$ for which $\sin\theta_0\approx\theta$.}
    \label{fig:SinTheta}
\end{figure}

Figure~\ref{fig:SinTheta} shows a plot of Eq.~(\ref{theta0vsX0}) as a function of $x_0$ and $l$. Also, the comparison between $\sin\theta_0$ and $\theta_0$ is presented. From the figure, it is clear that the approximation is valid for $\theta\leq 0.4$ rad which is satisfied with initial displacements $x_0<120$ cm and string lengths in the range $270\text{ cm}<l<300\text{ cm}$. Therefore, Eqs.~(\ref{T0}) and~(\ref{T1}) describe our experimental arrangements.
%--------------------------------------
%--------------------------------------

%%%%%%%%%%%%%%%%%%%%%%%%%%%%%%%%%%%%%%%%%%%
\subsection{Geographic dependence of the gravity acceleration}
%%%%%%%%%%%%%%%%%%%%%%%%%%%%%%%%%%%%%%%%%%%
In a standard university physics course, it is common to compute the acceleration $g$ of a body of mass $m$ falling by the gravitational attraction of a homogeneous sphere of radius $R$ and mass $M$. It leads to the formula:
\bea
g=G\frac{M}{R^2},
\label{gbasic}
\eea
where $G$ is the Cavendish constant~\cite{alonso1967fundamental,reese1998university}. By considering the values of the Earth's mass and equatorial radius, Eq.~(\ref{gbasic}) gives:
\bea
g\approx9.798~\text{m/s}^2,
\label{g0}
\eea
which is independent on the geographical coordinates.

%--------------------------------------
\begin{figure*}[t]
    \centering
    \includegraphics[scale=0.15]{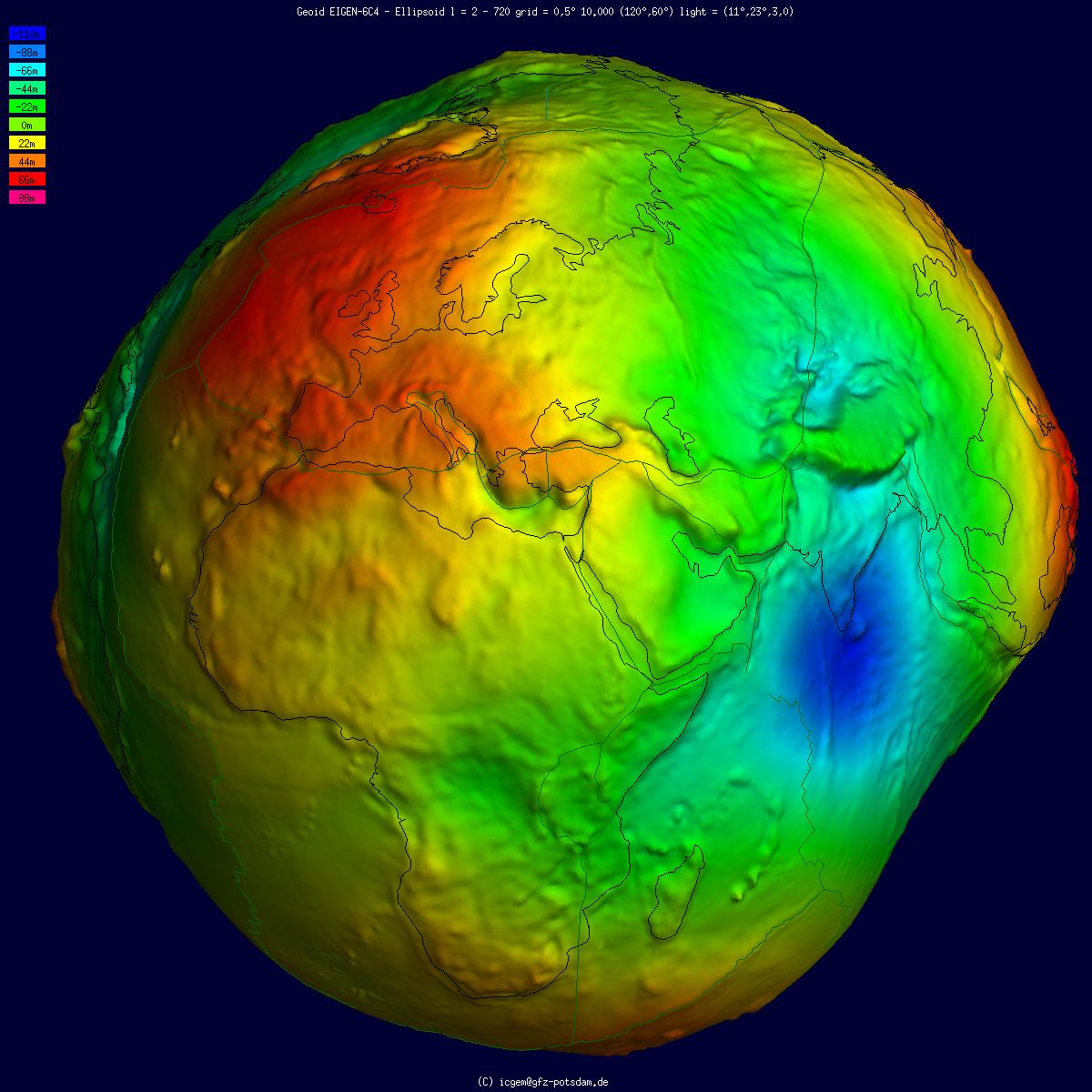}\hspace{0.6cm}\includegraphics[scale=0.15]{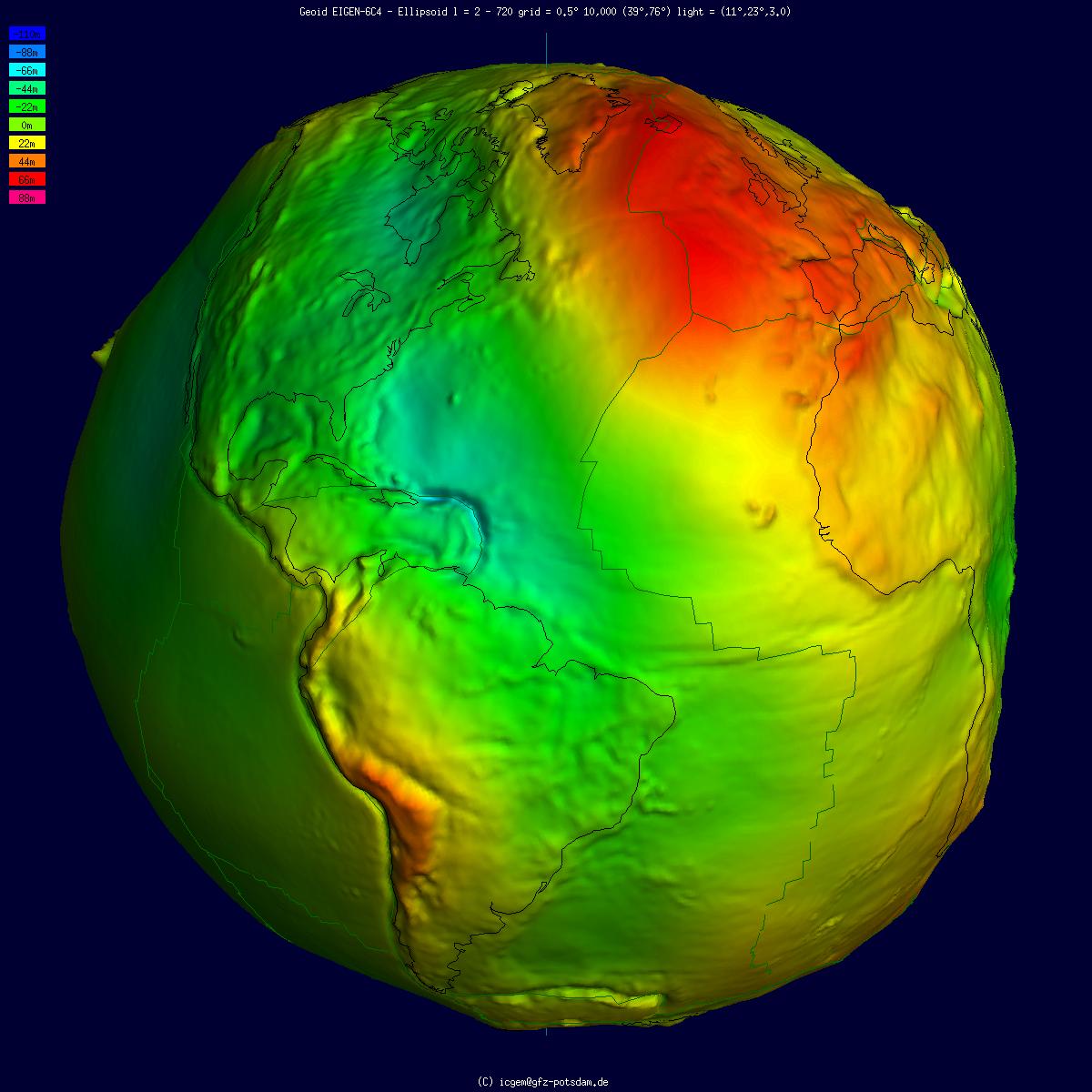}
    \caption{Geoid undulation in false color. The relief and vertical texture have been multiplied by a $\times 10^3$ factor. Images reproduced with the help of Ref.~\cite{ICGEMWeb}}
    \label{fig:geoid}
\end{figure*}
%--------------------------------------

It is well established that Earth is not a perfect sphere nor an ellipsoid. Moreover, it rotates and has local movements product of the tectonic evolution which implies deformations in its surface. So then, it is pretty evident that the gravitational acceleration changes locally. If the Earth's rotation is considered, the non-inertial frame produces a centrifugal acceleration. For a spherical shape, it leads to the formula~\cite{Gr_ber_2007}:
\bea
g_\text{eff}(\phi)=G\frac{M}{R^2}-\omega^2R\cos^2(\phi),
\eea
where $\omega$ is the angular frequency of rotation, and $\phi$ is the latitude coordinate measured in degrees. Then, if $R$ is taken as the Earth's equatorial radius, and $\omega=7.292 115\times10^{-5}\text{rad}/\text{s}$, the acceleration $g$ (in m$/$s$^2$) is given by:
\bea
g_\text{eff}(\phi)=9.798-0.034\cos^2(\phi).
\label{geff}
\eea

A more sophisticated description for the variations of $g$ with the latitude angle $\phi$ is addressed by assuming an anisotropic {mass distribution} by considering the Earth as a rotating ellipsoid~\cite{white1986world,hofmann2006physical,1978135}. For example, the following expression gives the latitude correction or the so-called {\it normal gravity}:
    \bea
    g_n(\phi)&=& 9.7803185[1 + 5.278895\times10^{-3} \sin^{2}(\phi)\nn\\
    &+&2.89\times10^{-5} \sin^{4}(\phi)], 
    \label{gn}
    \eea
where $\phi$ is measured in degrees, and $g$ in m$/$s$^2$. Nevertheless, this formula does not consider the variations of altitude, which is generally a very involved procedure given the non-homogeneous relief on the surface.

More realistically, some models of the shape of Earth are produced by measuring the gravity around the ``globe". For instance, a recent work computed the so-called {\it geoid} (an equipotential surface to which the force of gravity is everywhere perpendicular), which is shown in Fig.~\ref{fig:geoid}~\cite{ince2019icgem}. Although the shape and relief were exaggerated, it is possible to glimpse the non-uniformity of the gravitational field and the mass distribution over our planet. Of course, the WPA's equipment cannot measure the variation of the gravitational field with high precision, but as we show in Sec.~\ref{Computing $g$ as a function of the latitude, longitude, and altitude}, the experiments carried out with the network are sensitive to the altitude, longitude, and altitude.

%%%%%%%%%%%%%%%%%%%%%%%%%%%%%%%%%%%%%%%%%%%
\section{Description and process to Data acquisition WPA project UNAD}\label{sec:data_acquisition}

As commented in previous sections, the main objective of the WPA project is to form a network of remote experiments oriented to primary, secondary, and university physics education. The experiment consists of precise, remotely controlled pendulums whose design is geared towards maximum accuracy through a simple and replicable mechanical assembly. There are two pendulum models designed by the project: the primary pendulum WPA@ELAB and the secondary pendulum WPA@FREE, which differ in size, architecture, and design, being their mechanical parts similar. On the other hand, the universities associated with the project have a primary pendulum model and ten copies of the secondary pendulums. In this article, we will describe the experience of the pendulum network at UNAD.

The primary pendulum at UNAD (WPA@UNAD-BOG) was installed at the University's headquarters in Bogotá, D.C., Colombia. The geodetic coordinates of the final location of the project are latitude 4°35 N, longitude 74°51', and altitude of 2578 m above sea level, being the pendulum of the network located in a geographical area with higher altitude. Also, in the same city, there is another primary pendulum at Universidad de los Andes at $2500 \text{ m}$ altitude.

\begin{figure}[H]
    \centering
    \includegraphics[scale=0.3]{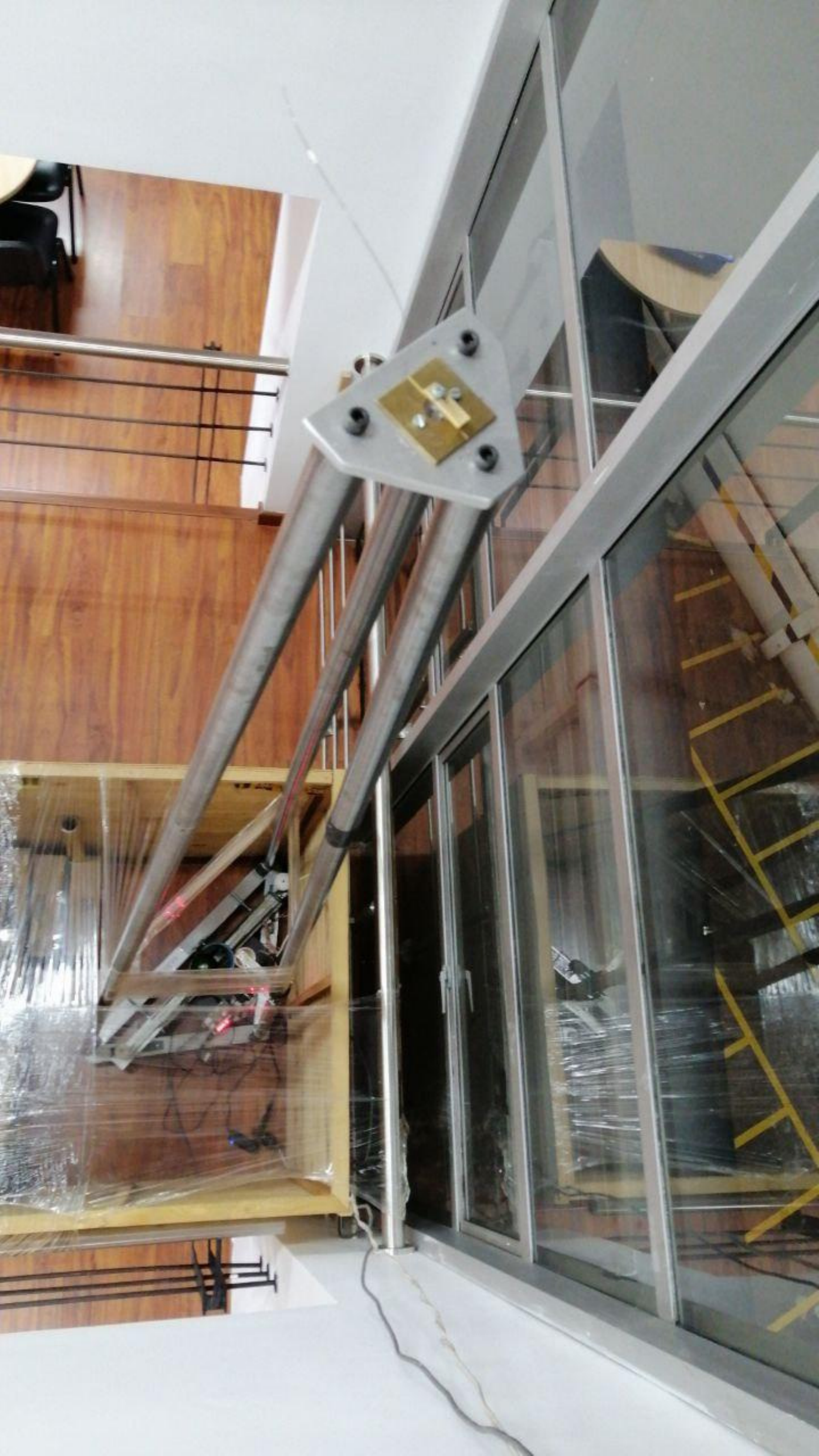}
    \caption{Photography of the Primary Pendulum  (WPA@UNAD)  installed at the main campus in Bogotá D.C. This pendulum together with the primaty pendulum of the University of the Andes are the highest elevation at sea level (2578 m above sea level), respect to the rest pendulums of the network.}
    \label{fig:pendulo_structure}
\end{figure}

The structure that supports the pendulum is a pyramidal metal structure of $2850 \pm 0,5 \text{ cm}$   in height. At the top of the pyramid has a trapezoidal fulcrum backed by metal support with a scratch that functions as a pivot. The mass of the pendulum is a 2 kg steel sphere held from the fulcrum by a round germanium wire of 0.45 mm thick and $2835\pm0.5\text{ cm}$ long. Figure~\ref{fig:pendulo_structure} shows a picture of the pendulum UNAD as installed in a fixed and isolated location.

The pendulum launcher consists of a paddle that will move the sphere previously fixed to the top of the structure. The movement is controlled by a motor that drives the direction of the rails. The details and cost for the pendulums installation can be found on the WPA web cite~\cite{WinNT}. Also, the interested reader can contact directly to the corresponding authors.

The software requirements are simple, given that this project is based on an open-access platform. The process of data acquisition and use of the pendulums network is explained in Ref.~\cite{ceur}. The operative system support Windows and Mac. There is no direct installation for OS GNU/Linux, but it is possible with involved processes. In the following briefly describes the use of the remote experiment, the entire installation process, and the activation of the experience (for extra details, consult Ref.~\cite{elab}).

First, the user must install the software associated with remote use and dependencies. The process of installing the experiment can be divided into three fundamental steps: 
\begin{enumerate}
    \item \textit{Install Java\textsuperscript{\texttrademark}}: The software that controls the movement of the pendulum is written in JAVA\textsuperscript{\texttrademark}. For this reason, it is necessary to install the latest version available on the official website~\cite{java}.
    
    \item \textit{Install VLC\textsuperscript{\texttrademark}}: VideoLan VLC\textsuperscript{\texttrademark} is the software associated with the video configuration, freely available. By installing it, you will be able to access the webcam of each experiment. It can be downloaded from the website~\cite{vlc}.
    
    \item \textit{Install e-Lab}: The e-lab software establishes the remote configuration of access to each of the remote experiments and access to the graphical user interface of the experiment controller. The installer is written in JAVA\textsuperscript{\texttrademark}, hence the need for initial installation before proceeding with the e-Lab. Install it from~\cite{elabweb}.
\end{enumerate}

There are four remote lab environments in the graphical user interface of the e-Lab software: \textit{basic}, \textit{intermediate}, \textit{advanced} and \textit{World Pendulum}, which can be found in English and Portuguese languages. To activate the pendulum network, it must be chosen from the list of experiments. A drop-down menu will show the network of active and inactive primary pendulums upon entering the interphase, so the user it can select an activate pendulum of the network. Once the user is in a interphase of a pendulum must to set up the horizontal launch offset and the number of samples to collect. The user can obtain the data of period, linear velocity and temperature with its corresponding uncertainties on the output screen. The graphical interface displays the period, velocity, and histogram plots ~\cite{ceur}. In the following section, we will use the WPA in order to compute the acceleration of gravity for several locations around the world.

%%%%%%%%%%%%%%%%%%%%%%%%%%%%%%%%%%%%%%%%%%%
\section{Results and discussion}\label{sec:Results_and_discussion}
%%%%%%%%%%%%%%%%%%%%%%%%%%%%%%%%%%%%%%%%%%%
This section shows and discusses the results of the remote experiments performed within the WPA. In particular, we analyze the impact of the number of repetitions and the initial amplitude, from which the calculation of the local gravitational acceleration is performed.

%%%%%%%%%%%%%%%%%%%%%%%%%%%%%%%%%%%%%%%%%%%
\subsection{The importance of repetitions}
%%%%%%%%%%%%%%%%%%%%%%%%%%%%%%%%%%%%%%%%%%%
%--------------------------------------------------
\begin{figure}[H]
    \centering
    \includegraphics[scale=0.5]{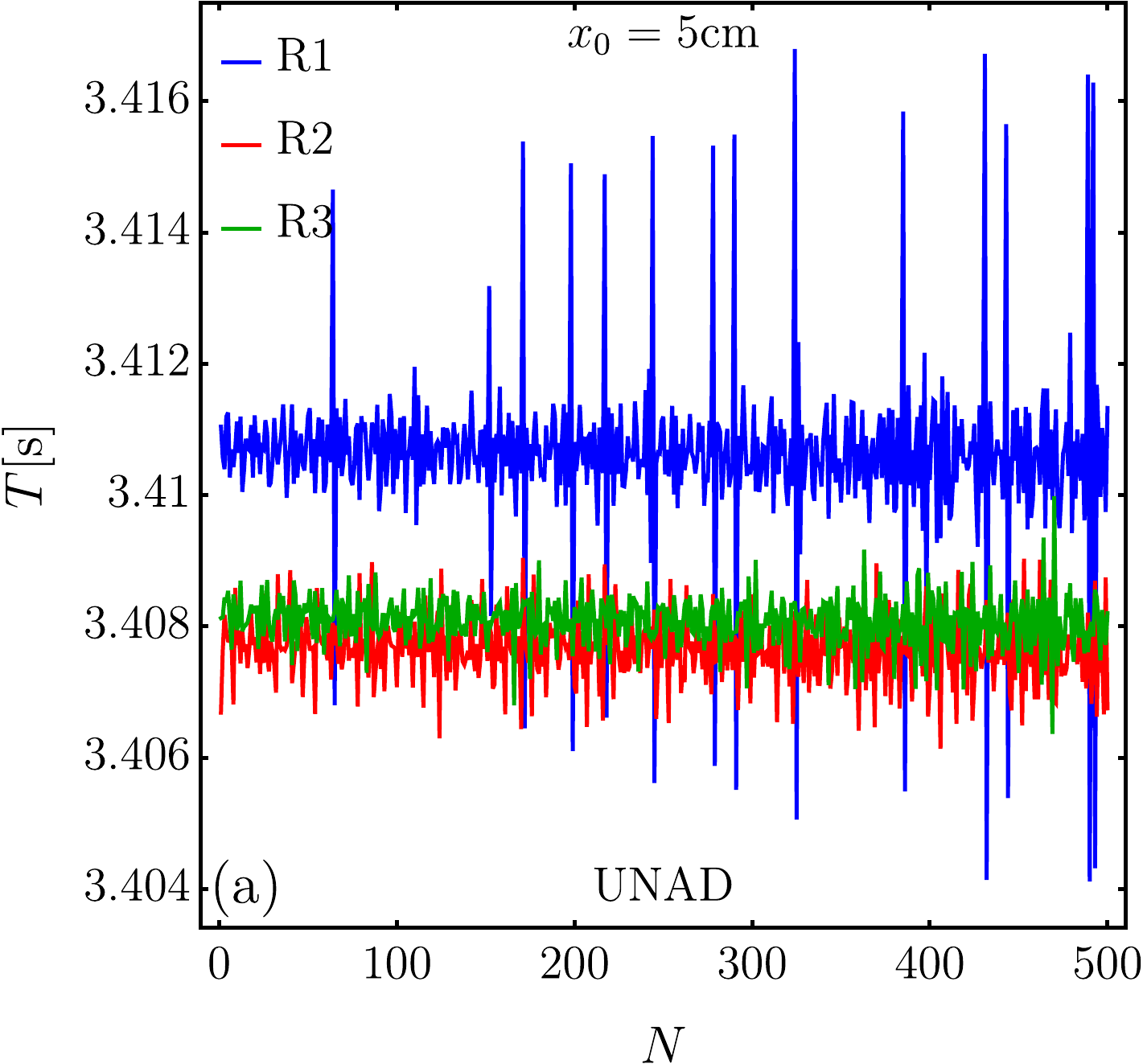}\\\vspace{0.5cm}
    \includegraphics[scale=0.5]{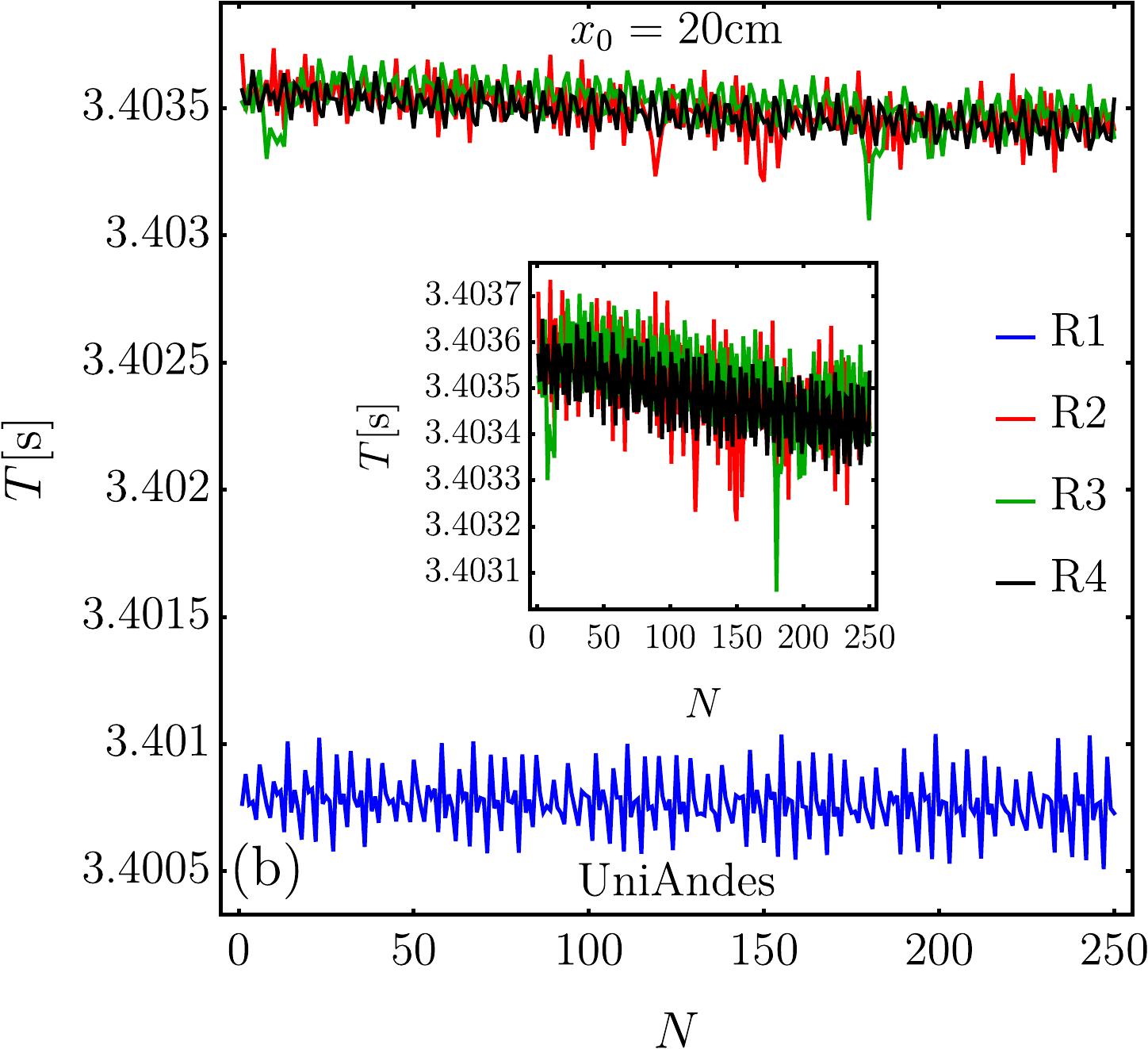}
    \caption{Period of the pendulum as a function of the number of samples $N$ for (a) $x_0=5$cm at UNAD, and (b) $x_0=20$cm at UniAndes. The color code indicates the set of data for each repetition of the experiment, and the inset in panel (b) is included to appreciate the scale.}
    \label{fig:role_of_R}
\end{figure}
%--------------------------------------------------

In an experiment, performing several repetitions is fundamental to achieving the best data set with a well-established confidence range. For example, figure.~\ref {fig:role_of_R} shows the measurements of the pendulum's period $T$ as a function of the number of samples $N$, for several repetitions (labeled as R1, R2, etc.), and for two initial amplitudes $x_0$. As can be noticed in Fig.~\ref{fig:role_of_R}(a), the first experiment (R1) has a considerable deviation from the data obtained in the other measurements (R2 and R3). A similar situation occurs in Fig.~\ref{fig:role_of_R}(b), where some data has a more pronounced shift in the vertical axis. Moreover, the experimental points are more dispersed when the number of samples grows. This behavior may be due to external perturbations like air-currents or synchronization problems. Nevertheless, it can be avoided by performing several repetitions in order to establish a pattern and realize when a group of data is valid or not. The latter illustrates one of the advantages of the present RCL: in a typical presential laboratory experience, data recollection is carried out in an established time window (usually the duration of a class). Therefore, the students do not have the opportunity to check in real-time if their data set meets the requirements of the experiment. In contrast, with the help of the WPA, the student can obtain samples at any time (provided by an Internet connection), which allows improving the analysis of the physics behind the phenomenon. 

% {\color{red}\bf On the other hand, note the decreasing behavior of the period in Fig.~\ref{fig:role_of_R}(b). In comparison with the data of Fig.~\ref{fig:role_of_R}(a), where the values of $T$ oscillate around a constant base-line, the case of $x_0=20$cm doesn't follow the model for a constant period. Then, given the value of the initial amplitude and the dimensions of the oscillating rod, one might assume that the sample shows forced oscillations???????.

% En la Fig.~\ref{fig:role_of_R}(b) el periodo disminuye, lo que quiere decir que lo están empujando. Al parecer también sucede con la Fig.~\ref{fig:UNAD_variosL} con $x_0=13$cm

% \begin{figure}
%     \centering
%     \includegraphics[scale=0.4]{UNAD_L20_N500-eps-converted-to.pdf}
%     \caption{El sistema parece forzado o los datos están mal}
%     \label{fig:my_label}
% \end{figure}

% }

%%%%%%%%%%%%%%%%%%%%%%%%%%%%%%%%%%%%%%%%%%%
\subsection{The role of $N$}
%%%%%%%%%%%%%%%%%%%%%%%%%%%%%%%%%%%%%%%%%%%
Another advantage of the RCL provided by the WPA is that it allows the user to perform $N$ samples for the pendulum's period. It is a significant feature in minimizing the error of the measurements. For instance, in an experiment where $N$-samples are taken, each with an error $\Delta x_n$, the error in the mean is given by~\cite{taylor1997introduction}:
\bea
\Delta x=N^{-1}\sqrt{\sum_{n=1}^N\Delta x_n^2},
\label{Deltax}
\eea
where the mean of $x$ is defined as:
\bea
\langle x\rangle\equiv N^{-1}\sum_{n=1}^N x_n.
\eea

%--------------------------------------------------
\begin{figure}[H]
    \centering
    \includegraphics[scale=0.55]{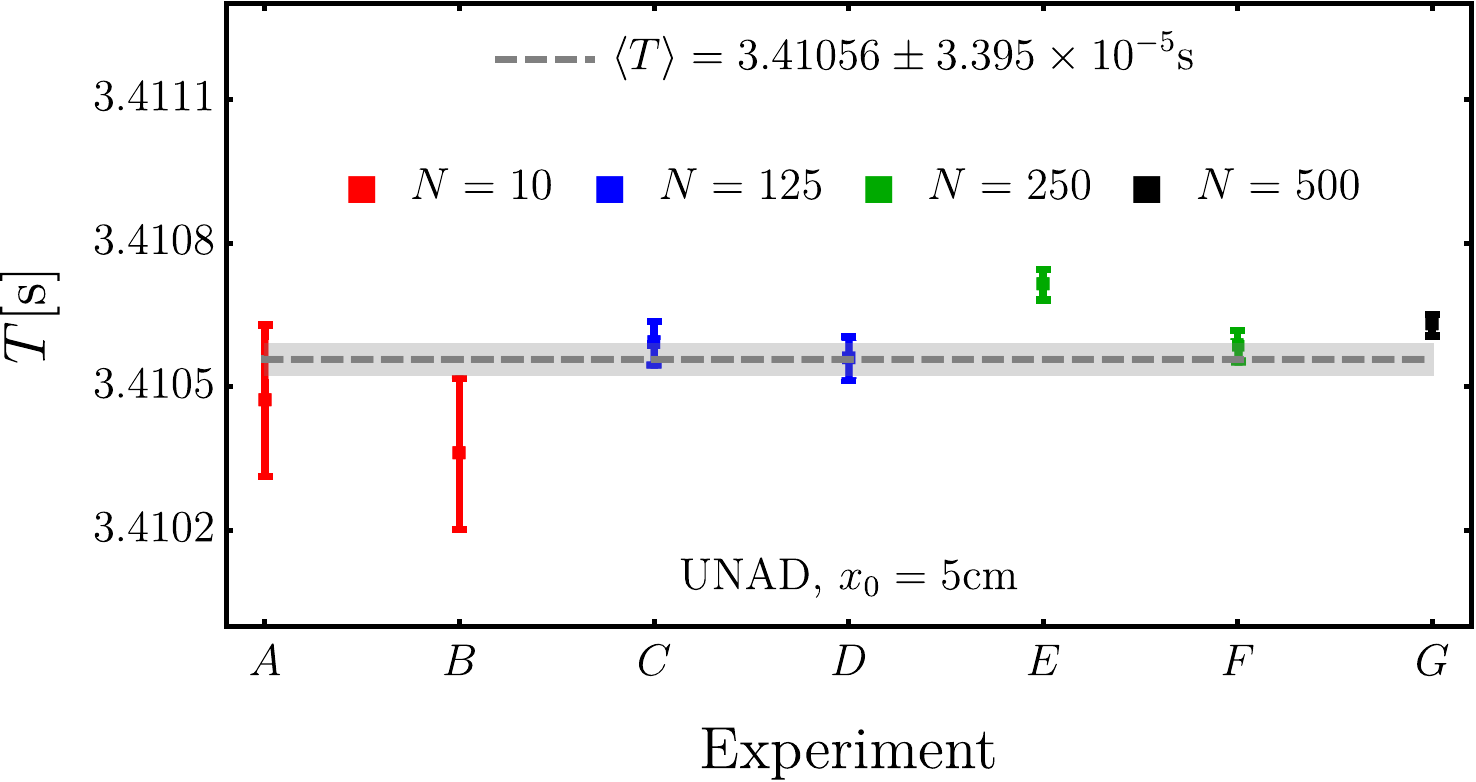}
    \caption{Period of the UNAD's pendulums with $x_0=5$cm for the experiments $A,B,C,D,E,F$ (see details in the text). The bars shows the error in the mean of each experiment. The dashed line and the gray region are the mean value over the experiments and its error, respectively.}
    \label{fig:TUNAD}
\end{figure}
%--------------------------------------------------

Note that from Eq.~(\ref{Deltax}), the error in the mean goes like $1/\sqrt{N}$ when the samples have the same $\Delta x$. So then, in order to ensure an accurate measure, a large $N$ must be considered. In that spirit, Fig.~\ref{fig:TUNAD} shows seven different experiments to measure the period $T$, with the UNAD's pendulum for the same initial configuration ($x_0=5$cm). The experiments $A$ and $B$ were performed with $N=10$ samples, $C$ and $D$ with $N=125$, $E$ and $F$ with $N=250$, and $G$ with $N=500$. In each experiment, the error is computed with Eq.~(\ref{Deltax}) and as was expected, it decreases when $N$ grows. Specifically, given that all the samples have the same experimental error, the error in the mean passes to be of order $\mathcal{O}(10^{-4}\text{s})$ for $N=10$ to $\mathcal{O}(10^{-5}\text{s})$ for $N=500$.

Depending on the particular analysis that the user wants to implement, several choices of data manipulation can be made. In particular, we take the data of the seven experiments as valid, and we perform the mean over them, which is shown in Fig.~\ref{fig:TUNAD} as a gray dashed line. The shaded region is its corresponding error, computed with Eq.~(\ref{Deltax}). In other words, we performed the mean over the means.

%%%%%%%%%%%%%%%%%%%%%%%%%%%%%%%%%%%%%%%%%%%
\subsection{Computing $g$ locally}
%%%%%%%%%%%%%%%%%%%%%%%%%%%%%%%%%%%%%%%%%%%
If we want to know some quantity $y$ that depends on the set of measured values $(x_1,\ldots,x_k)$, it is straightforward to find that the propagated error is~\cite{taylor1997introduction}:
\bea
\Delta y=\sqrt{\sum_{n=1}^k\left(\frac{\partial y}{\partial x_n} \right)^2\Delta x_n^2}.
\label{Deltay}
\eea

The above expression is used to compute the gravitational acceleration experimented by the UNAD's pendulum in the approximation given by Eq.~(\ref{T0}) so that: 
\begin{subequations}
\bea
g\approx\frac{4\pi^2L}{T^2},
\label{gg0}
\eea
and
\bea
\Delta g=g\sqrt{\frac{\Delta L^2}{L^2}+4\frac{\Delta T^2}{T^2}},
\label{errorg0}
\eea
\label{g0anderror}
\end{subequations}
where $L=l+R$ is the effective length of the pendulum.

%--------------------------------------------------
\begin{figure}[H]
    \centering
    \includegraphics[scale=0.57]{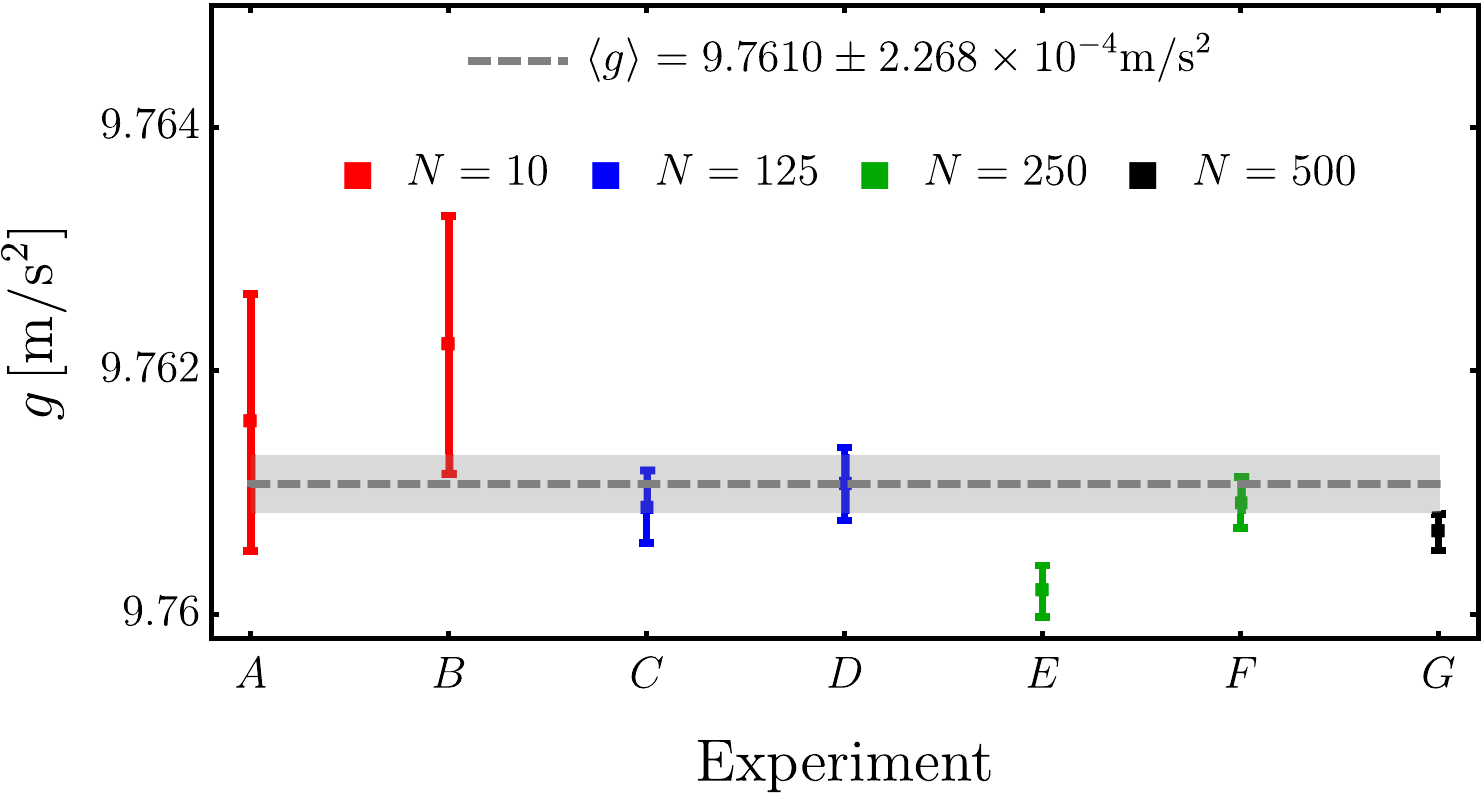}
    \caption{Gravitational acceleration $g$ experimented by the UNAD's pendulum computed with Eq.~(\ref{gg0}) for the same experiments of Fig.~\ref{fig:TUNAD}. The dashed line and the gray region are the mean value over the experiments and its error (computed with Eq.~(\ref{errorg0})), respectively.}
    \label{fig:gUNAD}
\end{figure}
%--------------------------------------------------

Figure~\ref{fig:gUNAD} shows the measurements of $g$ with the UNAD's pendulum for the discussed experiments. As can be noticed, the value of $\langle g\rangle$ is approximately the $99\%$ of the nominal acceleration of Eq.~(\ref{g0}). As we discuss in Sec.~\ref{Computing $g$ as a function of the latitude, longitude, and altitude}, it is due to Bogota's altitude, provided by the fact that $g$ in Eq.~(\ref{g0}) is assumed to be at the sea level.

%--------------------------------------------------
\begin{figure}[H]
    \centering
    \includegraphics[scale=0.55]{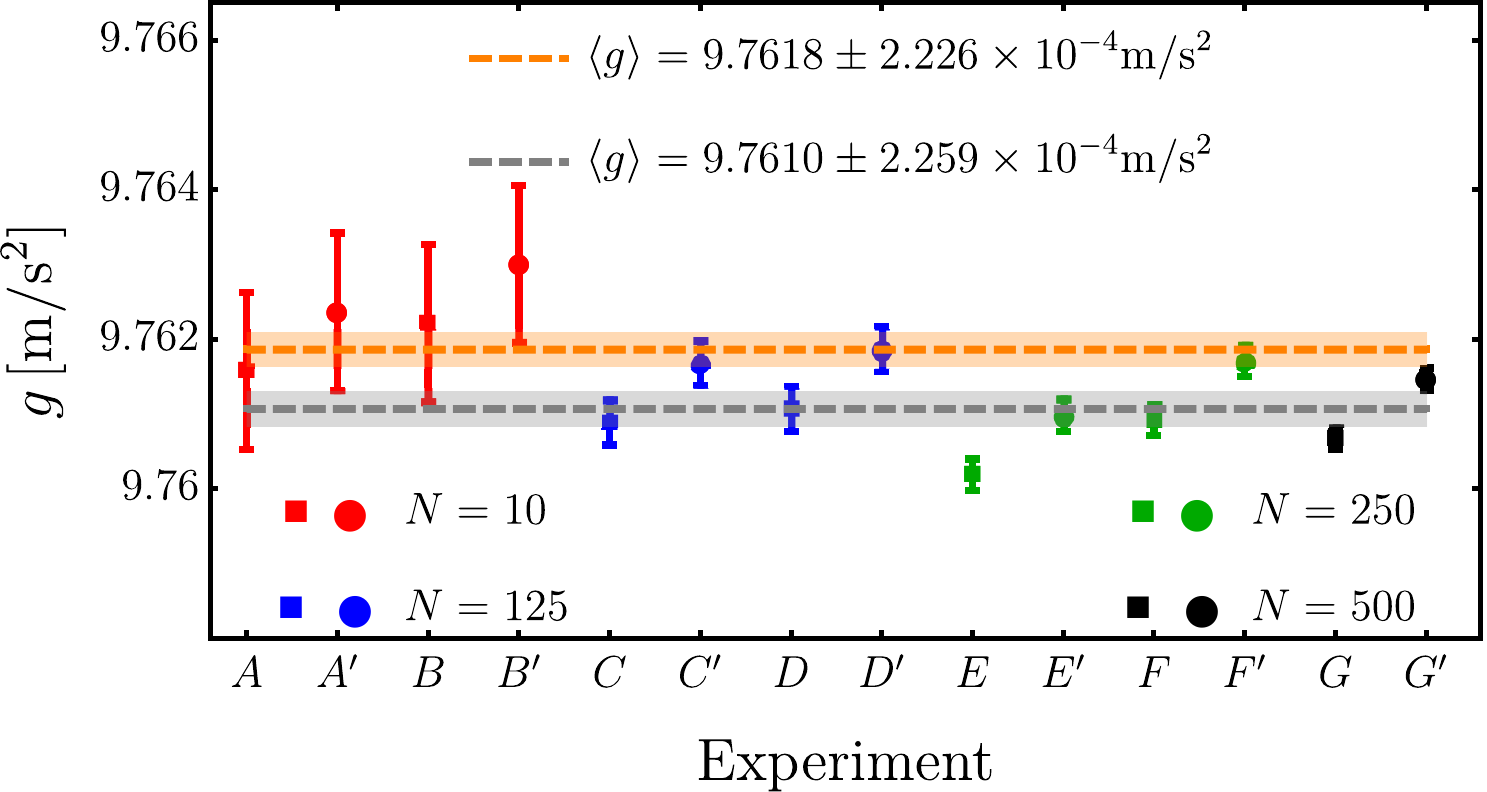}
    \caption{Comparison between the gravitational calculation performed with Eqs.~(\ref{g0anderror}) (squares with the gray dashed-line and region), and with Eqs.~(\ref{g1anderror}) (circles with primed labels and orange dashed-line/region).}
    \label{fig:comparison_inertia}
\end{figure}
%--------------------------------------------------

In order to improve the previous results, a second approximation is implemented by taking into account the moment of inertia of the pendulum's sphere. With the help of Eqs.~(\ref{T1}) and~(\ref{Deltay}) we get:
\begin{subequations}
\bea
g\approx\frac{4\pi^2}{5LT^2}(5L^2+2R^2),
\label{g1}
\eea
and
\bea
\Delta g&=&\frac{4\pi^2}{5LT^2}\Big[4(5L^2+2R^2)^2\frac{\Delta T^2}{T^2}+(5L^2-2R^2)^2\frac{\Delta L^2}{L^2}\nn\\
&+&16 R^2\Delta R^2\Big]^{1/2}.
\label{errorg1}
\eea
\label{g1anderror}
\end{subequations}

The comparison between the results of Eqs.~(\ref{g0anderror}) and Eqs.~(\ref{g1anderror}) is presented in Fig.~\ref{fig:comparison_inertia}. The value of $g$ with the moment of inertia contribution is just $0.8\%$ higher than the result obtained with the mathematical pendulum approximation. Of course, this is expected from Eq.~(\ref{T1}) given that:
\bea
\frac{2R^2}{5l^2}\Bigg|_\text{UNAD}\approx0.000083660.
\label{ratio}
\eea

Nevertheless, we include the corresponding analysis of the physical pendulum treatment for completeness and to give a perspective of its importance in precision measurements or scenarios where the ratio of Eq.~(\ref{ratio}) is a considerable quantity. In the following, given the parameters of Table~\ref{tab:ExperimentalParameters} such a contribution is ignored. 

%%%%%%%%%%%%%%%%%%%%%%%%%%%%%%%%%%%%%%%%%%%
\subsection{Computing $g$ as a function of the latitude, longitude, and altitude}\label{Computing $g$ as a function of the latitude, longitude, and altitude}
%%%%%%%%%%%%%%%%%%%%%%%%%%%%%%%%%%%%%%%%%%%

\begin{figure}[H]
    \centering
    \includegraphics[scale=0.55]{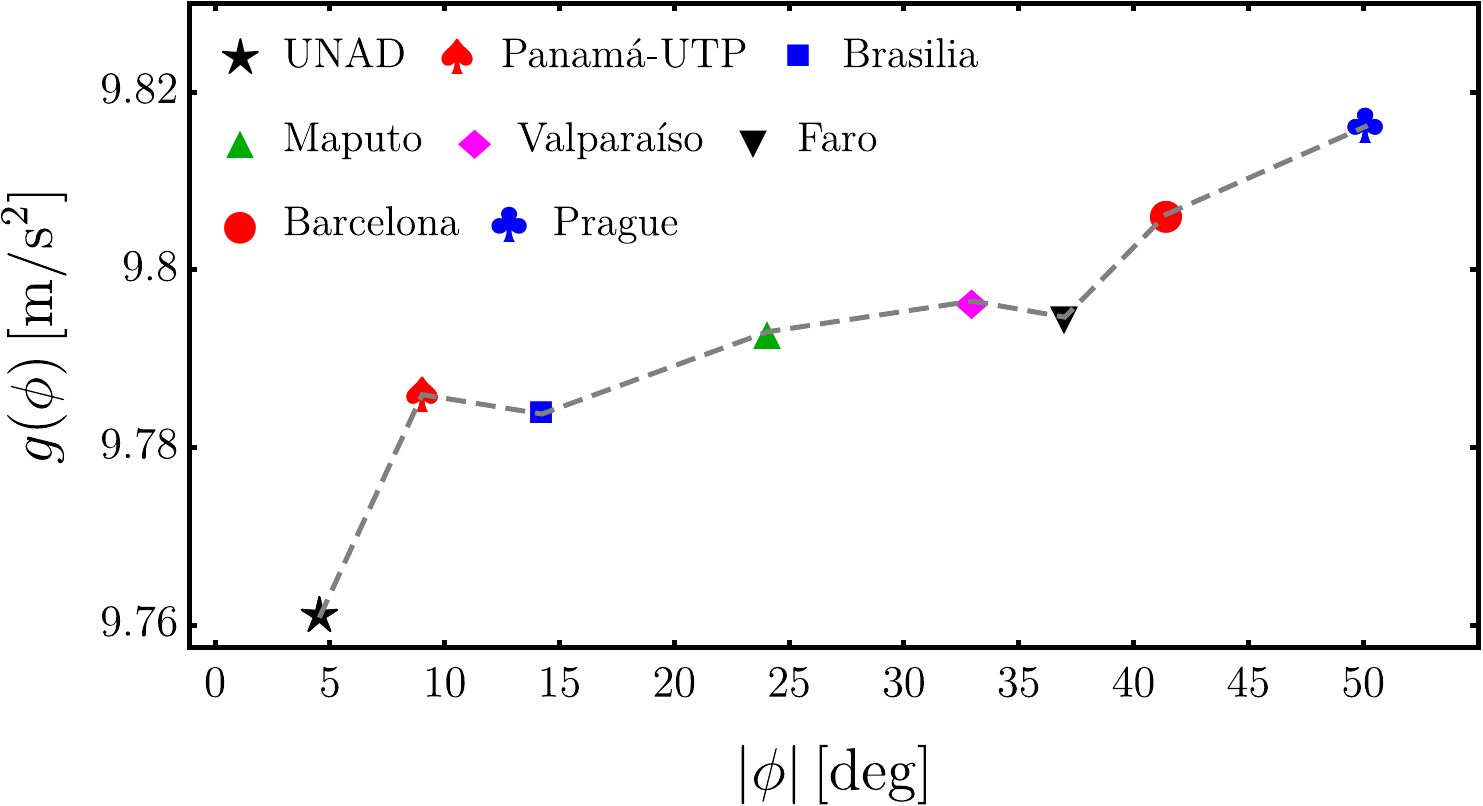}
    \caption{Gravitational acceleration computed from Eq.~(\ref{gg0}) as a function of the latitude $\phi$ for several pendulums in the WPA (see Table~\ref{tab:ExperimentalParameters}). Only the central value is shown, given the small magnitude of the error bars.}
    \label{fig:g_phi}
\end{figure}

One of the attractive features in the WPA is the fact that the remote experiment can be performed in several countries, which allows studying the variation of gravity with the geographic location. By following the methodology discussed in the previous section, Fig.~\ref{fig:g_phi} shows the value of $g$ as a function of the latitude $\phi$. As a consequence of the approximately ellipsoidal form of our planet, the gravitational acceleration increases with latitude and has slight variations around the nominal value $g_0$.
%--------------------------------------------------
%--------------------------------------------------

\begin{figure}[H]
    \centering
    \includegraphics[scale=0.55]{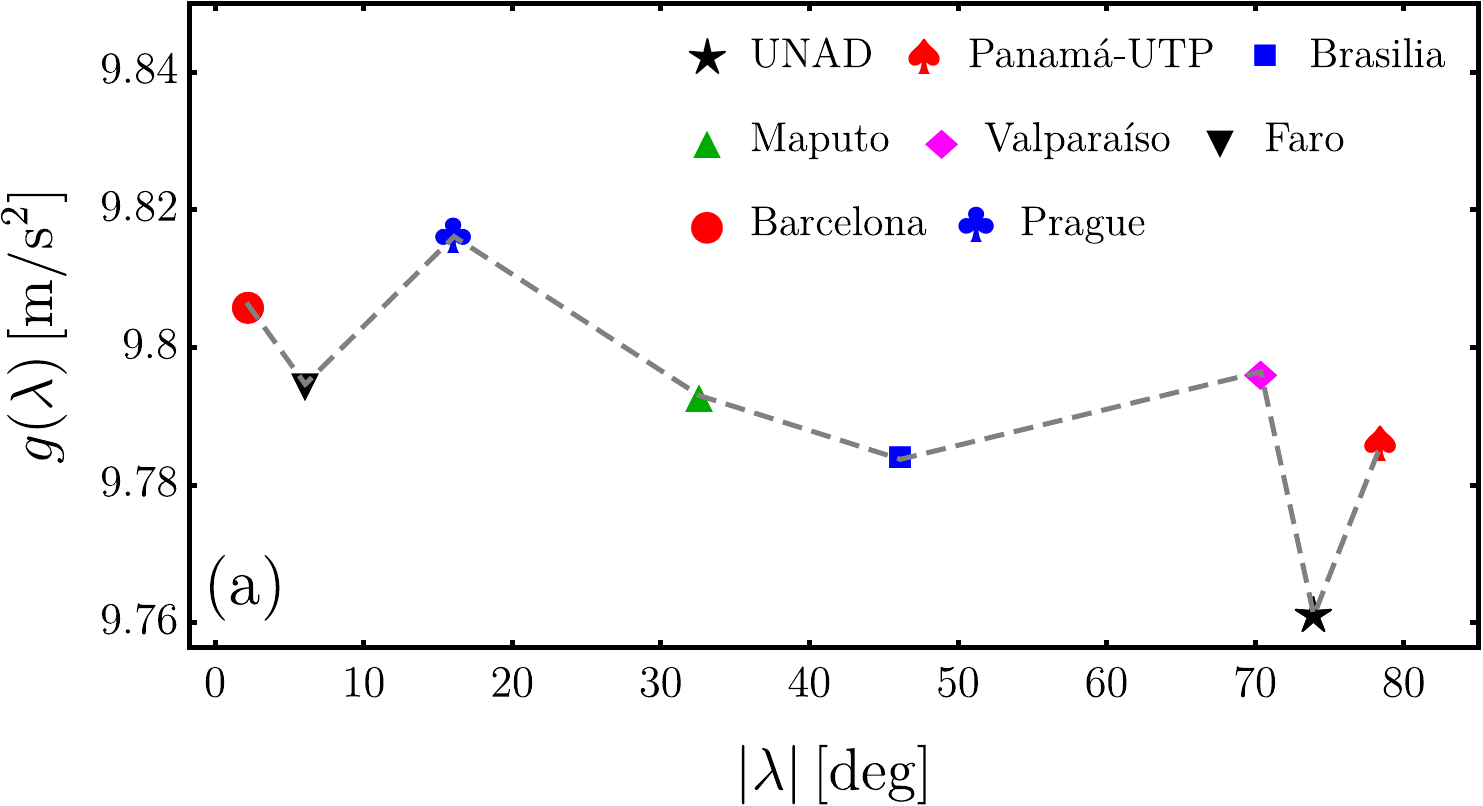}\\
    \hspace{0.3cm}
    \includegraphics[scale=0.55]{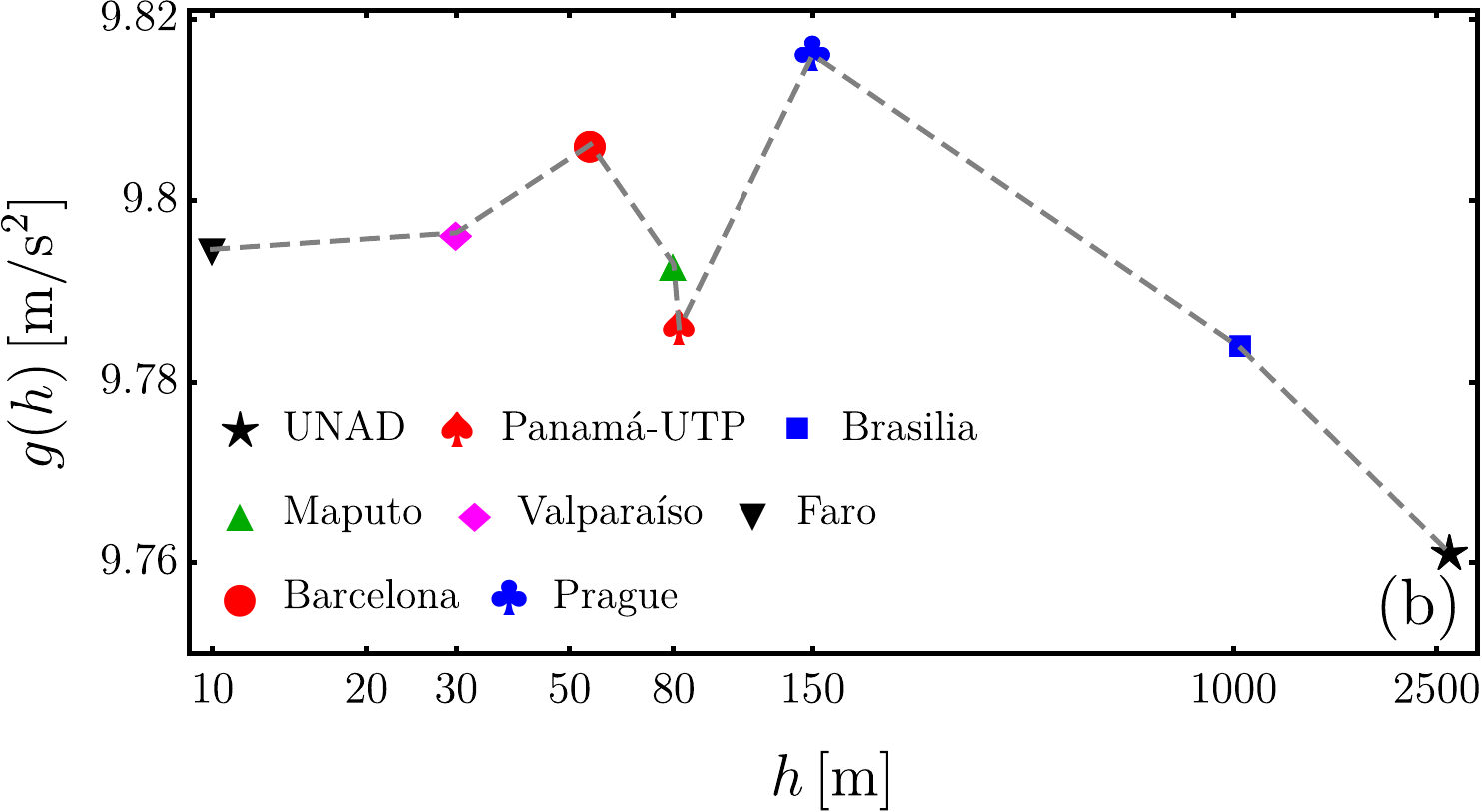}
    \caption{Gravitational acceleration computed from Eq.~(\ref{gg0}) as a function of (a) the longitude $\lambda$, and (b) the altitude $h$, for several pendulums in the WPA (see Table~\ref{tab:ExperimentalParameters}). Only the central value is shown, given the small magnitude of the error bars.}
    \label{fig:g_lambda_h}
\end{figure}

For completeness, the variations of $g$ with respect to the longitude $\lambda$, and the altitude $h$ are shown in Figs.~\ref{fig:g_lambda_h}(a) and~\ref{fig:g_lambda_h}(b), respectively. Note that the local gravitational field is a non-trivial function of the coordinates $(\phi,\lambda,h)$.

%--------------------------------------------------
%--------------------------------------------------
For instance, one may assume that the points corresponding to Maputo and Panama-UTP would have a minor $g$ than Prague (because of their altitude), but it turns out to be a combination of $\phi$ and $\lambda$ such that the latter city results with the highest local gravity in the intervals considered. This is an interesting effect to be presented to students in order to discuss the planet's mass distribution.

\begin{figure}[H]
    \centering
    \includegraphics[scale=0.55]{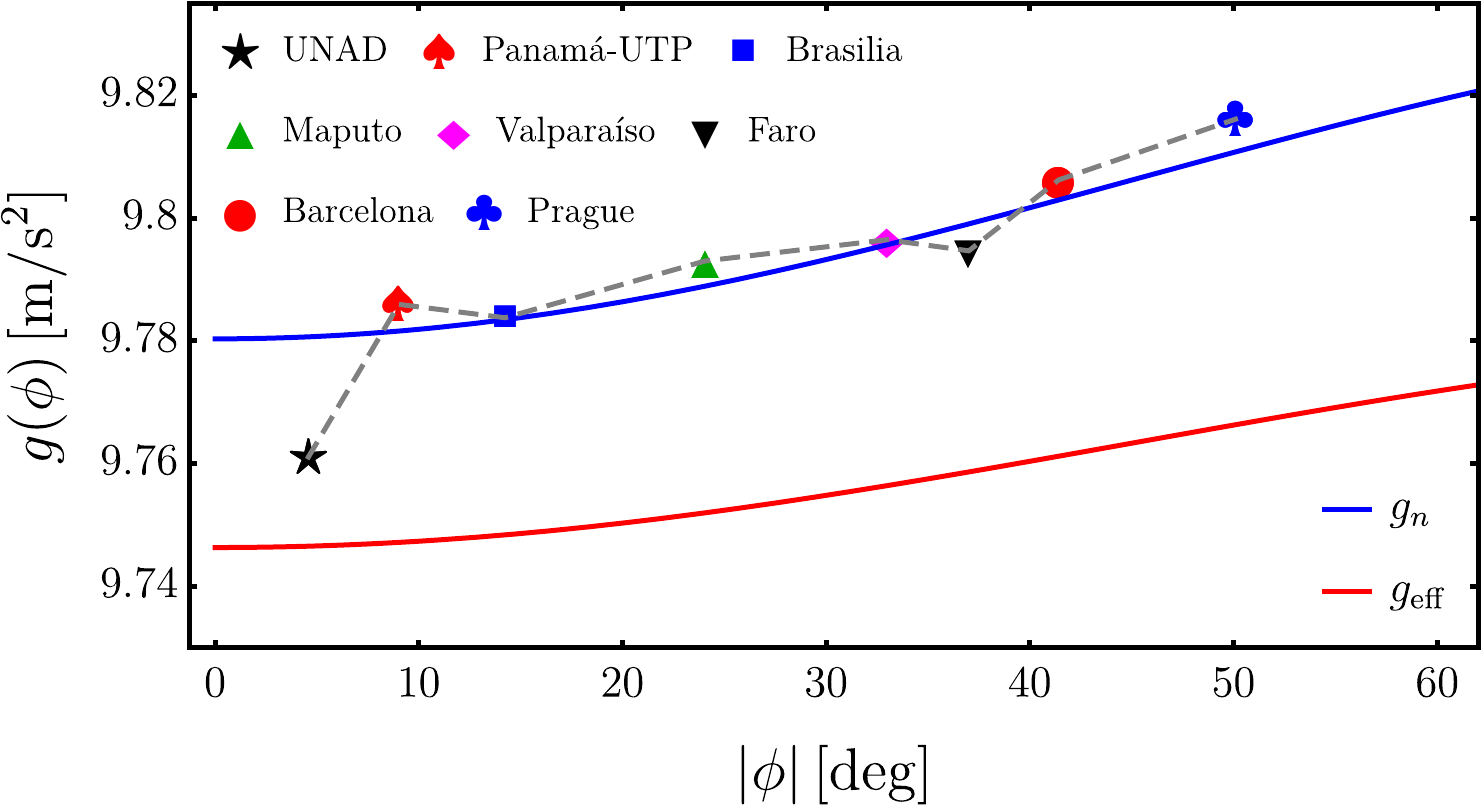}
    \caption{Comparison between the experimental data and the models of Eqs.~(\ref{geff}) and~(\ref{gn}) for the latitude-dependent gravitational acceleration. Only the central value is shown, given the small magnitude of the error bars.}
    \label{fig:g_comparative}
\end{figure}

\begin{figure}[H]
    \centering
    \includegraphics[scale=0.15]{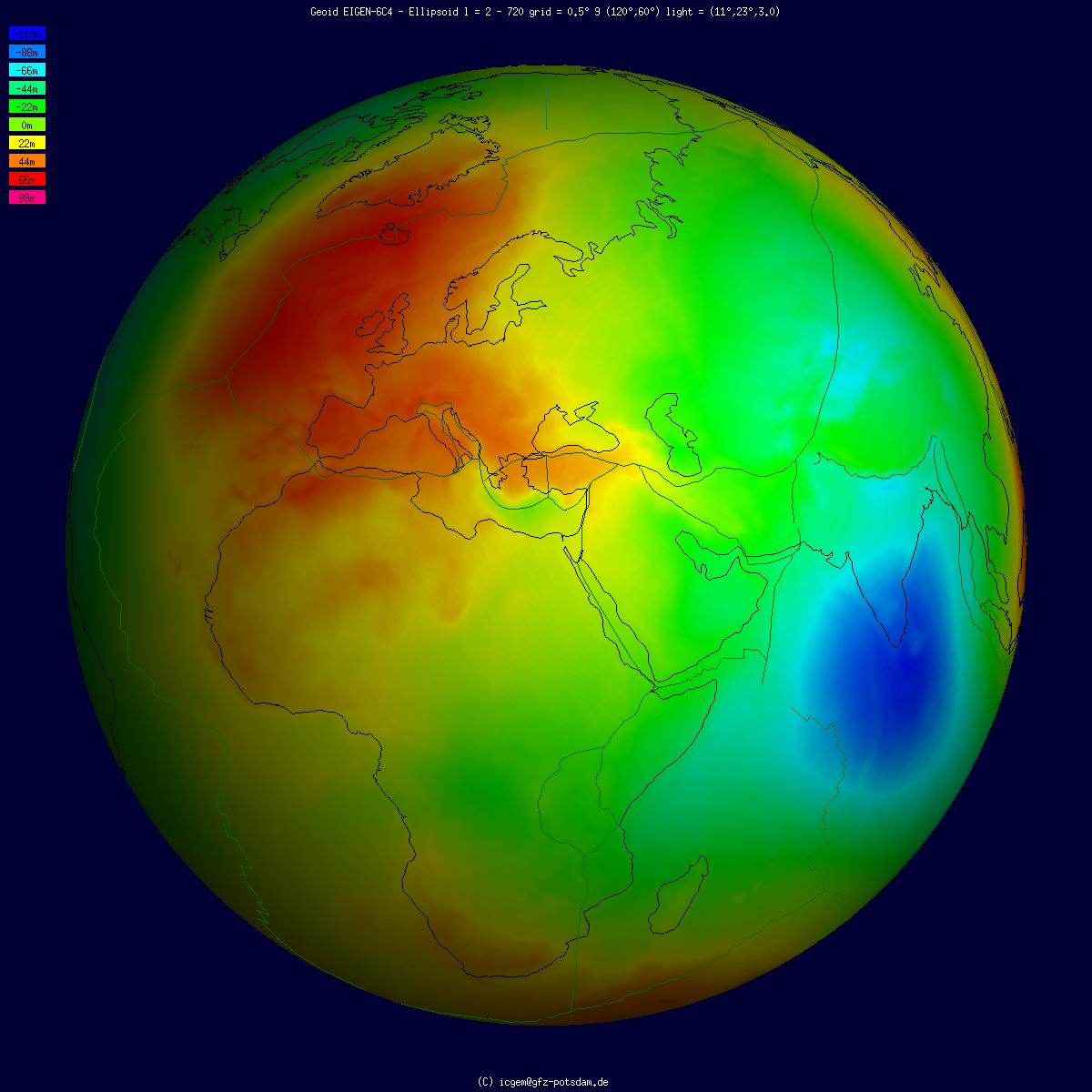}
    \caption{Geoid undulation in false color. The relief and vertical texture haven't been multiplied by an scale factor. Image reproduced with the help of Ref.~\cite{ICGEMWeb}}
    \label{fig:geoid3}
\end{figure}

Finally, our results can give an image of the average Earth's shape. For this sake, in Fig.~\ref{fig:g_comparative} the experimental data and the models of Eqs.~(\ref{geff}) and~(\ref{gn}) for the latitude-dependent gravitational acceleration are shown. The results fit the analytical expression for a rotating ellipsoid, and the fluctuations around it may come from the changes in altitude (the model considers the gravity at sea level). Nevertheless, the expression for $g_n$ assumes for the ellipsoid an equatorial radius $R_\text{eq}\approx 6378.137$ km, and a polar radius $R_\text{pol}\approx6356.752$ km~\cite{Gr_ber_2007}. The latter, combined with the geoid without scale factor (used in Fig.~\ref{fig:geoid} to appreciate the relief), leads to the picture of the Earth very close to a sphere, as is shown in Fig.~\ref{fig:geoid3}.

%--------------------------------------------------
% \subsection{Next phase of the WPA: Installation of the remote secondary pendulums in UNAD.}

% Section IV discussed the WPA@FREE secondary pendulum network, an evolution of the WPA@ELAB project, started with the installation of remote laboratories at each partner institution, the purpose of which will be to manage an even larger global network of remote experiments.

% \begin{figure}[H]
%     \centering
%     \includegraphics[scale=0.65]{map.png}
%     \caption{}
%     \label{fig:pendulum_map}
% \end{figure}

%--------------------------------------------------
%--------------------------------------------------

%--------------------------------------
%--------------------------------------
%%%%%%%%%%%%%%%%%%%%%%%%%%%%%%%%%%%%%
\section{Summary and Conclusions}\label{sec:Summary and Conclusions}
%%%%%%%%%%%%%%%%%%%%%%%%%%%%%%%%%%%%%
In this work, we have presented the WPA as a network for remote access experiments to bring additional educational tools into the frame of physics teaching. We discussed the main features of the locations, installation, and data acquisition for the online experience. The primary and secondary pendulums are used to obtain experimental data to compute the local gravitational acceleration from a fundamental physics framework. The analysis of the pendulum's parameters allows us to work in a small amplitude approximation. Then, the value of $g$ is determined by the length and period of each setup. The latter is corroborated when the experimental data is contrasted with the rigid solid expression. 

To obtain a well-estimated gravity acceleration, we emphasize the importance of repetitions in a single experience. The latter is because several sources of error can be found, which are minimized when the sample is large enough. The results show that the period of the pendulums in the network is indeed sensitive to latitude, longitude, and altitude. In fact, the computed value of $g$ is compared with theoretical/phenomenological models for its latitude-dependence with an excellent agreement of the experimental data.

Moreover, deviations from the analytical expressions are interpreted in terms of the altitude and longitude of each location, given that the named equations assume perfect spherical or ellipsoidal symmetries. The latter is extrapolated to an interpretation of the Earth's shape. With the model's restrictions, we can conclude that our planet is almost spherical, and deviations from that form are imperceptible with single experiments. 

We hope this tool and the presented analysis will be helpful for introductory and university physics courses. Also, given the new paradigms in educational strategies, we encourage the educating community to embrace these new technologies and tools in order to improve its presentations and classes.

%%%%%%%%%%%%%%%%%%%%%%%%%%%%%%%%%%%%%
\section*{CRediT author statement}
%%%%%%%%%%%%%%%%%%%%%%%%%%%%%%%%%%%%%
\textbf{Freddy Torres-Payoma:} Software, Validation, Investigation, Resources, Data Curation,  Writing – original draft, Writing - Review \& Editing, Project administration, Funding acquisition.

\textbf{Diana Herrera:}  Writing – original draft, Writing - Review \& Editing, Project administration, Funding acquisition.

\textbf{Karla Triana:}  Writing – original draft, Writing - Review \& Editing, Project administration, Funding acquisition.

\textbf{Laura Neira-Quintero:}  Validation, Investigation, Data Curation,  Writing – original draft.

\textbf{Jorge David Casta\~no-Yepes:} Conceptualization, Methodology, Software, Validation, Formal analysis, Investigation, Data Curation, Writing - Original Draft, Writing - Review \& Editing, Visualization, Supervision.\\

%%%%%%%%%%%%%%%%%%%%%%%%%%%%%%%%%%%%%
\section*{Declaration of Competing interests}
%%%%%%%%%%%%%%%%%%%%%%%%%%%%%%%%%%%%%
The authors declare that they have no known competing financial interests or personal relationships that could have appeared to influence the work reported in this paper.

%%%%%%%%%%%%%%%%%%%%%%%%%%%%%%%%%%%%%
\section*{Acknowledgements}
%%%%%%%%%%%%%%%%%%%%%%%%%%%%%%%%%%%%%
J.D.C.-Y. thanks Erika J. Moreno and Andrés Medina for their valuable comments and suggested references. The authors thank the European Commission for the grant {\it Erasmus+ Project WP@ELAB 598929-EPP-1-PT-EPPKA2-CBHE-JP World Pendulum Alliance}. Authos thank to Melina Castaño for a thorough reading of the manuscript and the language correction.

\raggedright
\bibliography{WPA.bib}

\begin{thebibliography}{35}%
\makeatletter
\providecommand \@ifxundefined [1]{%
 \@ifx{#1\undefined}
}%
\providecommand \@ifnum [1]{%
 \ifnum #1\expandafter \@firstoftwo
 \else \expandafter \@secondoftwo
 \fi
}%
\providecommand \@ifx [1]{%
 \ifx #1\expandafter \@firstoftwo
 \else \expandafter \@secondoftwo
 \fi
}%
\providecommand \natexlab [1]{#1}%
\providecommand \enquote  [1]{``#1''}%
\providecommand \bibnamefont  [1]{#1}%
\providecommand \bibfnamefont [1]{#1}%
\providecommand \citenamefont [1]{#1}%
\providecommand \href@noop [0]{\@secondoftwo}%
\providecommand \href [0]{\begingroup \@sanitize@url \@href}%
\providecommand \@href[1]{\@@startlink{#1}\@@href}%
\providecommand \@@href[1]{\endgroup#1\@@endlink}%
\providecommand \@sanitize@url [0]{\catcode `\\12\catcode `\$12\catcode
  `\&12\catcode `\#12\catcode `\^12\catcode `\_12\catcode `\%12\relax}%
\providecommand \@@startlink[1]{}%
\providecommand \@@endlink[0]{}%
\providecommand \url  [0]{\begingroup\@sanitize@url \@url }%
\providecommand \@url [1]{\endgroup\@href {#1}{\urlprefix }}%
\providecommand \urlprefix  [0]{URL }%
\providecommand \Eprint [0]{\href }%
\providecommand \doibase [0]{http://dx.doi.org/}%
\providecommand \selectlanguage [0]{\@gobble}%
\providecommand \bibinfo  [0]{\@secondoftwo}%
\providecommand \bibfield  [0]{\@secondoftwo}%
\providecommand \translation [1]{[#1]}%
\providecommand \BibitemOpen [0]{}%
\providecommand \bibitemStop [0]{}%
\providecommand \bibitemNoStop [0]{.\EOS\space}%
\providecommand \EOS [0]{\spacefactor3000\relax}%
\providecommand \BibitemShut  [1]{\csname bibitem#1\endcsname}%
\let\auto@bib@innerbib\@empty
%</preamble>
\bibitem [{\citenamefont {Miller}\ \emph {et~al.}(2015)\citenamefont {Miller},
  \citenamefont {Schell}, \citenamefont {Ho}, \citenamefont {Lukoff},\ and\
  \citenamefont {Mazur}}]{PhysRevSTPER.11.010104}%
  \BibitemOpen
  \bibfield  {author} {\bibinfo {author} {\bibfnamefont {Kelly}\ \bibnamefont
  {Miller}}, \bibinfo {author} {\bibfnamefont {Julie}\ \bibnamefont {Schell}},
  \bibinfo {author} {\bibfnamefont {Andrew}\ \bibnamefont {Ho}}, \bibinfo
  {author} {\bibfnamefont {Brian}\ \bibnamefont {Lukoff}}, \ and\ \bibinfo
  {author} {\bibfnamefont {Eric}\ \bibnamefont {Mazur}},\ }\bibfield  {title}
  {\enquote {\bibinfo {title} {Response switching and self-efficacy in peer
  instruction classrooms},}\ }\href {\doibase 10.1103/PhysRevSTPER.11.010104}
  {\bibfield  {journal} {\bibinfo  {journal} {Phys. Rev. ST Phys. Educ. Res.}\
  }\textbf {\bibinfo {volume} {11}},\ \bibinfo {pages} {010104} (\bibinfo
  {year} {2015})}\BibitemShut {NoStop}%
\bibitem [{\citenamefont {Kortemeyer}(2007)}]{PhysRevSTPER.3.010101}%
  \BibitemOpen
  \bibfield  {author} {\bibinfo {author} {\bibfnamefont {Gerd}\ \bibnamefont
  {Kortemeyer}},\ }\bibfield  {title} {\enquote {\bibinfo {title} {Correlations
  between student discussion behavior, attitudes, and learning},}\ }\href
  {\doibase 10.1103/PhysRevSTPER.3.010101} {\bibfield  {journal} {\bibinfo
  {journal} {Phys. Rev. ST Phys. Educ. Res.}\ }\textbf {\bibinfo {volume}
  {3}},\ \bibinfo {pages} {010101} (\bibinfo {year} {2007})}\BibitemShut
  {NoStop}%
\bibitem [{\citenamefont {Kirstein}\ and\ \citenamefont
  {Nordmeier}(2007)}]{Kirstein_2007}%
  \BibitemOpen
  \bibfield  {author} {\bibinfo {author} {\bibfnamefont {Juergen}\ \bibnamefont
  {Kirstein}}\ and\ \bibinfo {author} {\bibfnamefont {Volkhard}\ \bibnamefont
  {Nordmeier}},\ }\bibfield  {title} {\enquote {\bibinfo {title} {Multimedia
  representation of experiments in physics},}\ }\href {\doibase
  10.1088/0143-0807/28/3/s11} {\bibfield  {journal} {\bibinfo  {journal} {Eur.
  J. Phys.}\ }\textbf {\bibinfo {volume} {28}},\ \bibinfo {pages} {S115--S126}
  (\bibinfo {year} {2007})}\BibitemShut {NoStop}%
\bibitem [{\citenamefont {Wagner}\ \emph {et~al.}(2006)\citenamefont {Wagner},
  \citenamefont {Altherr}, \citenamefont {Eckert},\ and\ \citenamefont
  {Jodl}}]{Wagner_2006}%
  \BibitemOpen
  \bibfield  {author} {\bibinfo {author} {\bibfnamefont {Andreas}\ \bibnamefont
  {Wagner}}, \bibinfo {author} {\bibfnamefont {Stefan}\ \bibnamefont
  {Altherr}}, \bibinfo {author} {\bibfnamefont {Bodo}\ \bibnamefont {Eckert}},
  \ and\ \bibinfo {author} {\bibfnamefont {Hans~Jörg}\ \bibnamefont {Jodl}},\
  }\bibfield  {title} {\enquote {\bibinfo {title} {Multimedia in physics
  education: two teaching videos on the absorption and emission spectrum of
  sodium},}\ }\href {\doibase 10.1088/0143-0807/27/6/l01} {\bibfield  {journal}
  {\bibinfo  {journal} {Eur. J. Phys.}\ }\textbf {\bibinfo {volume} {27}},\
  \bibinfo {pages} {L31--L35} (\bibinfo {year} {2006})}\BibitemShut {NoStop}%
\bibitem [{\citenamefont {Gröber}\ \emph {et~al.}(2007)\citenamefont
  {Gröber}, \citenamefont {Vetter}, \citenamefont {Eckert},\ and\
  \citenamefont {Jodl}}]{Gr_ber_2007}%
  \BibitemOpen
  \bibfield  {author} {\bibinfo {author} {\bibfnamefont {S}~\bibnamefont
  {Gröber}}, \bibinfo {author} {\bibfnamefont {M}~\bibnamefont {Vetter}},
  \bibinfo {author} {\bibfnamefont {B}~\bibnamefont {Eckert}}, \ and\ \bibinfo
  {author} {\bibfnamefont {H-J}\ \bibnamefont {Jodl}},\ }\bibfield  {title}
  {\enquote {\bibinfo {title} {{World pendulum{\textemdash}a distributed
  remotely controlled laboratory ({RCL}) to measure the Earth's gravitational
  acceleration depending on geographical latitude}},}\ }\href {\doibase
  10.1088/0143-0807/28/3/022} {\bibfield  {journal} {\bibinfo  {journal} {Eur.
  J. Phys.}\ }\textbf {\bibinfo {volume} {28}},\ \bibinfo {pages} {603--613}
  (\bibinfo {year} {2007})}\BibitemShut {NoStop}%
\bibitem [{\citenamefont {Afanador}(2021)}]{afanador2021educacion}%
  \BibitemOpen
  \bibfield  {author} {\bibinfo {author} {\bibfnamefont {Jaime Alberto~Leal}\
  \bibnamefont {Afanador}},\ }\enquote {\bibinfo {title} {{Educaci{\'o}n,
  virtualidad e innovaci{\'o}n: Estudio de caso para la consolidaci{\'o}n de un
  modelo de liderazgo en la educaci{\'o}n incluyente y de calidad}},}\ \
  (\bibinfo  {publisher} {Sello Editorial UNAD},\ \bibinfo {year} {2021})\
  \bibinfo {edition} {1st}\ ed.\BibitemShut {Stop}%
\bibitem [{\citenamefont {Torres}\ \emph {et~al.}(2019)\citenamefont {Torres},
  \citenamefont {Gonçalves}, \citenamefont {Ricardo}, \citenamefont
  {Ferreira}, \citenamefont {Calado}, \citenamefont {Torres}, \citenamefont
  {Fernandes},\ and\ \citenamefont {Guerra}}]{torres2019collaborative}%
  \BibitemOpen
  \bibfield  {author} {\bibinfo {author} {\bibfnamefont {André}\ \bibnamefont
  {Torres}}, \bibinfo {author} {\bibfnamefont {Duarte}\ \bibnamefont
  {Gonçalves}}, \bibinfo {author} {\bibfnamefont {Emanuel}\ \bibnamefont
  {Ricardo}}, \bibinfo {author} {\bibfnamefont {Ricardo~Grosso}\ \bibnamefont
  {Ferreira}}, \bibinfo {author} {\bibfnamefont {Rui}\ \bibnamefont {Calado}},
  \bibinfo {author} {\bibfnamefont {Rui}\ \bibnamefont {Torres}}, \bibinfo
  {author} {\bibfnamefont {Horácio}\ \bibnamefont {Fernandes}}, \ and\
  \bibinfo {author} {\bibfnamefont {Vasco}\ \bibnamefont {Guerra}},\ }\bibfield
   {title} {\enquote {\bibinfo {title} {{Collaborative development of plasma
  physics MOOC in the context of a PhD curricular unit}},}\ }in\ \href
  {\doibase 10.1109/EXPAT.2019.8876481} {\emph {\bibinfo {booktitle} {2019 5th
  Experiment International Conference (exp.at'19)}}}\ (\bibinfo {year} {2019})\
  pp.\ \bibinfo {pages} {123--127}\BibitemShut {NoStop}%
\bibitem [{\citenamefont {Drinkwater}\ \emph {et~al.}(2003)\citenamefont
  {Drinkwater}, \citenamefont {Floberghagen}, \citenamefont {Haagmans},
  \citenamefont {Muzi},\ and\ \citenamefont {Popescu}}]{Drinkwater2003}%
  \BibitemOpen
  \bibfield  {author} {\bibinfo {author} {\bibfnamefont {M.~R.}\ \bibnamefont
  {Drinkwater}}, \bibinfo {author} {\bibfnamefont {R.}~\bibnamefont
  {Floberghagen}}, \bibinfo {author} {\bibfnamefont {R.}~\bibnamefont
  {Haagmans}}, \bibinfo {author} {\bibfnamefont {D.}~\bibnamefont {Muzi}}, \
  and\ \bibinfo {author} {\bibfnamefont {A.}~\bibnamefont {Popescu}},\ }in\
  \href {\doibase 10.1007/978-94-017-1333-7_36} {\emph {\bibinfo {booktitle}
  {{GOCE: ESA's First Earth Explorer Core Mission}}}},\ \bibinfo {editor}
  {edited by\ \bibinfo {editor} {\bibfnamefont {G.}~\bibnamefont {Beutler}},
  \bibinfo {editor} {\bibfnamefont {M.~R.}\ \bibnamefont {Drinkwater}},
  \bibinfo {editor} {\bibfnamefont {R.}~\bibnamefont {Rummel}}, \ and\ \bibinfo
  {editor} {\bibfnamefont {R.}~\bibnamefont {Von~Steiger}}}\ (\bibinfo
  {publisher} {Springer Netherlands},\ \bibinfo {address} {Dordrecht},\
  \bibinfo {year} {2003})\ pp.\ \bibinfo {pages} {419--432}\BibitemShut
  {NoStop}%
\bibitem [{\citenamefont {Rebhan}\ \emph {et~al.}(2000)\citenamefont {Rebhan},
  \citenamefont {Aguirre},\ and\ \citenamefont
  {Johannessen}}]{rebhan2000gravity}%
  \BibitemOpen
  \bibfield  {author} {\bibinfo {author} {\bibfnamefont {H}~\bibnamefont
  {Rebhan}}, \bibinfo {author} {\bibfnamefont {M}~\bibnamefont {Aguirre}}, \
  and\ \bibinfo {author} {\bibfnamefont {J}~\bibnamefont {Johannessen}},\
  }\bibfield  {title} {\enquote {\bibinfo {title} {{The gravity field and
  steady-state ocean circulation explorer mission-GOCE}},}\ }\href
  {https://earth.esa.int/goce04/Documents/eoq66_goce.pdf} {\bibfield  {journal}
  {\bibinfo  {journal} {Earth Obs. Q.}\ }\textbf {\bibinfo {volume} {66}},\
  \bibinfo {pages} {6--11} (\bibinfo {year} {2000})}\BibitemShut {NoStop}%
\bibitem [{\citenamefont {Sampietro}(2016)}]{Sampietro2016}%
  \BibitemOpen
  \bibfield  {author} {\bibinfo {author} {\bibfnamefont {Daniele}\ \bibnamefont
  {Sampietro}},\ }\bibfield  {title} {\enquote {\bibinfo {title} {Crustal
  modelling and moho estimation with goce gravity data},}\ }in\ \href
  {https://doi.org/10.1007/978-3-319-16952-1_8} {\emph {\bibinfo {booktitle}
  {Remote sensing advances for earth system science}}}\ (\bibinfo  {publisher}
  {Springer},\ \bibinfo {year} {2016})\ pp.\ \bibinfo {pages}
  {127--144}\BibitemShut {NoStop}%
\bibitem [{\citenamefont {Kornfeld}\ \emph {et~al.}(2019)\citenamefont
  {Kornfeld}, \citenamefont {Arnold}, \citenamefont {Gross}, \citenamefont
  {Dahya}, \citenamefont {Klipstein}, \citenamefont {Gath},\ and\ \citenamefont
  {Bettadpur}}]{GRACE3}%
  \BibitemOpen
  \bibfield  {author} {\bibinfo {author} {\bibfnamefont {Richard~P.}\
  \bibnamefont {Kornfeld}}, \bibinfo {author} {\bibfnamefont {Bradford~W.}\
  \bibnamefont {Arnold}}, \bibinfo {author} {\bibfnamefont {Michael~A.}\
  \bibnamefont {Gross}}, \bibinfo {author} {\bibfnamefont {Neil~T.}\
  \bibnamefont {Dahya}}, \bibinfo {author} {\bibfnamefont {William~M.}\
  \bibnamefont {Klipstein}}, \bibinfo {author} {\bibfnamefont {Peter~F.}\
  \bibnamefont {Gath}}, \ and\ \bibinfo {author} {\bibfnamefont {Srinivas}\
  \bibnamefont {Bettadpur}},\ }\bibfield  {title} {\enquote {\bibinfo {title}
  {{GRACE-FO: The Gravity Recovery and Climate Experiment Follow-On
  Mission}},}\ }\href {\doibase 10.2514/1.A34326} {\bibfield  {journal}
  {\bibinfo  {journal} {J. Spacecr Rockets}\ }\textbf {\bibinfo {volume}
  {56}},\ \bibinfo {pages} {931--951} (\bibinfo {year} {2019})}\BibitemShut
  {NoStop}%
\bibitem [{\citenamefont {Swenson}\ and\ \citenamefont {Wahr}(2002)}]{GRACE1}%
  \BibitemOpen
  \bibfield  {author} {\bibinfo {author} {\bibfnamefont {Sean}\ \bibnamefont
  {Swenson}}\ and\ \bibinfo {author} {\bibfnamefont {John}\ \bibnamefont
  {Wahr}},\ }\bibfield  {title} {\enquote {\bibinfo {title} {{Methods for
  inferring regional surface-mass anomalies from Gravity Recovery and Climate
  Experiment (GRACE) measurements of time-variable gravity}},}\ }\href
  {\doibase https://doi.org/10.1029/2001JB000576} {\bibfield  {journal}
  {\bibinfo  {journal} {J. Geophys. Res.}\ }\textbf {\bibinfo {volume} {107}},\
  \bibinfo {pages} {ETG 3--1--ETG 3--13} (\bibinfo {year} {2002})}\BibitemShut
  {NoStop}%
\bibitem [{\citenamefont {Niu}\ \emph {et~al.}(2007)\citenamefont {Niu},
  \citenamefont {Yang}, \citenamefont {Dickinson}, \citenamefont {Gulden},\
  and\ \citenamefont {Su}}]{GRACE2}%
  \BibitemOpen
  \bibfield  {author} {\bibinfo {author} {\bibfnamefont {Guo-Yue}\ \bibnamefont
  {Niu}}, \bibinfo {author} {\bibfnamefont {Zong-Liang}\ \bibnamefont {Yang}},
  \bibinfo {author} {\bibfnamefont {Robert~E.}\ \bibnamefont {Dickinson}},
  \bibinfo {author} {\bibfnamefont {Lindsey~E.}\ \bibnamefont {Gulden}}, \ and\
  \bibinfo {author} {\bibfnamefont {Hua}\ \bibnamefont {Su}},\ }\bibfield
  {title} {\enquote {\bibinfo {title} {{Development of a simple groundwater
  model for use in climate models and evaluation with Gravity Recovery and
  Climate Experiment data}},}\ }\href {\doibase
  https://doi.org/10.1029/2006JD007522} {\bibfield  {journal} {\bibinfo
  {journal} {J. Geophys. Res.}\ }\textbf {\bibinfo {volume} {112}} (\bibinfo
  {year} {2007}),\ https://doi.org/10.1029/2006JD007522}\BibitemShut {NoStop}%
\bibitem [{\citenamefont {Chambers}\ \emph {et~al.}(2004)\citenamefont
  {Chambers}, \citenamefont {Wahr},\ and\ \citenamefont {Nerem}}]{GRACE4}%
  \BibitemOpen
  \bibfield  {author} {\bibinfo {author} {\bibfnamefont {Don~P.}\ \bibnamefont
  {Chambers}}, \bibinfo {author} {\bibfnamefont {John}\ \bibnamefont {Wahr}}, \
  and\ \bibinfo {author} {\bibfnamefont {R.~Steven}\ \bibnamefont {Nerem}},\
  }\bibfield  {title} {\enquote {\bibinfo {title} {{Preliminary observations of
  global ocean mass variations with GRACE}},}\ }\href {\doibase
  https://doi.org/10.1029/2004GL020461} {\bibfield  {journal} {\bibinfo
  {journal} {Geophys. Res. Lett.}\ }\textbf {\bibinfo {volume} {31}} (\bibinfo
  {year} {2004}),\ https://doi.org/10.1029/2004GL020461}\BibitemShut {NoStop}%
\bibitem [{\citenamefont {Turcotte}\ and\ \citenamefont
  {Schubert}(2002)}]{turcotte2002geodynamics}%
  \BibitemOpen
  \bibfield  {author} {\bibinfo {author} {\bibfnamefont {Donald~L}\
  \bibnamefont {Turcotte}}\ and\ \bibinfo {author} {\bibfnamefont {Gerald}\
  \bibnamefont {Schubert}},\ }\enquote {\bibinfo {title} {Geodynamics},}\ \
  (\bibinfo  {publisher} {Cambridge university press},\ \bibinfo {year}
  {2002})\BibitemShut {NoStop}%
\bibitem [{\citenamefont {Chen}\ \emph {et~al.}(2019)\citenamefont {Chen},
  \citenamefont {Wu},\ and\ \citenamefont {Suppe}}]{chen2019southward}%
  \BibitemOpen
  \bibfield  {author} {\bibinfo {author} {\bibfnamefont {Yi-Wei}\ \bibnamefont
  {Chen}}, \bibinfo {author} {\bibfnamefont {Jonny}\ \bibnamefont {Wu}}, \ and\
  \bibinfo {author} {\bibfnamefont {John}\ \bibnamefont {Suppe}},\ }\bibfield
  {title} {\enquote {\bibinfo {title} {{Southward propagation of Nazca
  subduction along the Andes}},}\ }\href
  {https://www.nature.com/articles/s41586-018-0860-1} {\bibfield  {journal}
  {\bibinfo  {journal} {Nature}\ }\textbf {\bibinfo {volume} {565}},\ \bibinfo
  {pages} {441--447} (\bibinfo {year} {2019})}\BibitemShut {NoStop}%
\bibitem [{\citenamefont {Boonma}\ \emph {et~al.}(2019)\citenamefont {Boonma},
  \citenamefont {Kumar}, \citenamefont {Garc{\'\i}a-Castellanos}, \citenamefont
  {Jim{\'e}nez-Munt},\ and\ \citenamefont
  {Fern{\'a}ndez}}]{boonma2019lithospheric}%
  \BibitemOpen
  \bibfield  {author} {\bibinfo {author} {\bibfnamefont {Kittiphon}\
  \bibnamefont {Boonma}}, \bibinfo {author} {\bibfnamefont {Ajay}\ \bibnamefont
  {Kumar}}, \bibinfo {author} {\bibfnamefont {Daniel}\ \bibnamefont
  {Garc{\'\i}a-Castellanos}}, \bibinfo {author} {\bibfnamefont {Ivone}\
  \bibnamefont {Jim{\'e}nez-Munt}}, \ and\ \bibinfo {author} {\bibfnamefont
  {Manel}\ \bibnamefont {Fern{\'a}ndez}},\ }\bibfield  {title} {\enquote
  {\bibinfo {title} {Lithospheric mantle buoyancy: the role of tectonic
  convergence and mantle composition},}\ }\href
  {https://www.nature.com/articles/s41598-019-54374-w} {\bibfield  {journal}
  {\bibinfo  {journal} {Sci. Rep.}\ }\textbf {\bibinfo {volume} {9}},\ \bibinfo
  {pages} {1--8} (\bibinfo {year} {2019})}\BibitemShut {NoStop}%
\bibitem [{\citenamefont {Moreno}\ and\ \citenamefont
  {Manea}(2021)}]{MORENO2021103604}%
  \BibitemOpen
  \bibfield  {author} {\bibinfo {author} {\bibfnamefont {Erika~Jessenia}\
  \bibnamefont {Moreno}}\ and\ \bibinfo {author} {\bibfnamefont {Marina}\
  \bibnamefont {Manea}},\ }\bibfield  {title} {\enquote {\bibinfo {title}
  {{Geodynamic evaluation of the pacific tectonic model for chortis block
  evolution using 3D numerical models of subduction}},}\ }\href {\doibase
  https://doi.org/10.1016/j.jsames.2021.103604} {\bibfield  {journal} {\bibinfo
   {journal} {J. S. Am. Earth Sci.}\ }\textbf {\bibinfo {volume} {112}},\
  \bibinfo {pages} {103604} (\bibinfo {year} {2021})}\BibitemShut {NoStop}%
\bibitem [{\citenamefont {Dumberry}\ and\ \citenamefont
  {Bloxham}(2004)}]{dumberry2004variations}%
  \BibitemOpen
  \bibfield  {author} {\bibinfo {author} {\bibfnamefont {Mathieu}\ \bibnamefont
  {Dumberry}}\ and\ \bibinfo {author} {\bibfnamefont {Jeremy}\ \bibnamefont
  {Bloxham}},\ }\bibfield  {title} {\enquote {\bibinfo {title} {Variations in
  the earth's gravity field caused by torsional oscillations in the core},}\
  }\href {https://academic.oup.com/gji/article/159/2/417/2064436} {\bibfield
  {journal} {\bibinfo  {journal} {Geophys. J. Int.}\ }\textbf {\bibinfo
  {volume} {159}},\ \bibinfo {pages} {417--434} (\bibinfo {year}
  {2004})}\BibitemShut {NoStop}%
\bibitem [{\citenamefont {Escobar}\ \emph {et~al.}(2019)\citenamefont
  {Escobar}, \citenamefont {Fernandes}, \citenamefont {Allard},\ and\
  \citenamefont {Erazo}}]{escobar2019pendulum}%
  \BibitemOpen
  \bibfield  {author} {\bibinfo {author} {\bibfnamefont {Manuel}\ \bibnamefont
  {Escobar}}, \bibinfo {author} {\bibfnamefont {Hor{\'a}cio}\ \bibnamefont
  {Fernandes}}, \bibinfo {author} {\bibfnamefont {Orlando}\ \bibnamefont
  {Allard}}, \ and\ \bibinfo {author} {\bibfnamefont {Yeni}\ \bibnamefont
  {Erazo}},\ }\bibfield  {title} {\enquote {\bibinfo {title} {Pendulum as an
  educational remote experiment},}\ }in\ \href
  {https://ieeexplore.ieee.org/abstract/document/8876473} {\emph {\bibinfo
  {booktitle} {2019 5th Experiment International Conference (exp. at'19)}}}\
  (\bibinfo {organization} {IEEE},\ \bibinfo {year} {2019})\ pp.\ \bibinfo
  {pages} {388--393}\BibitemShut {NoStop}%
\bibitem [{Win(2021)}]{WinNT}%
  \BibitemOpen
  \href {http://www.elab.ist.utl.pt/wiki/index.php?title=P%C3%A9ndulo_mundial}
  {\enquote {\bibinfo {title} {Péndulo mundial},}\ } (\bibinfo {year}
  {2021})\BibitemShut {NoStop}%
\bibitem [{\citenamefont {Landau}(1973)}]{landau1973classical}%
  \BibitemOpen
  \bibfield  {author} {\bibinfo {author} {\bibfnamefont {Lev~D}\ \bibnamefont
  {Landau}},\ }\enquote {\bibinfo {title} {Classical mechanics},}\ \ (\bibinfo
  {publisher} {Nauka},\ \bibinfo {year} {1973})\BibitemShut {NoStop}%
\bibitem [{\citenamefont {Alonso}\ and\ \citenamefont
  {Finn}(1967)}]{alonso1967fundamental}%
  \BibitemOpen
  \bibfield  {author} {\bibinfo {author} {\bibfnamefont {Marcelo}\ \bibnamefont
  {Alonso}}\ and\ \bibinfo {author} {\bibfnamefont {Edward~J}\ \bibnamefont
  {Finn}},\ }\enquote {\bibinfo {title} {Fundamental university physics},}\ \
  (\bibinfo  {publisher} {Addison-Wesley Reading, MA},\ \bibinfo {year}
  {1967})\BibitemShut {NoStop}%
\bibitem [{\citenamefont {Reese}(1998)}]{reese1998university}%
  \BibitemOpen
  \bibfield  {author} {\bibinfo {author} {\bibfnamefont {Ronald~Lane}\
  \bibnamefont {Reese}},\ }\enquote {\bibinfo {title} {University physics},}\ \
  (\bibinfo  {publisher} {Brooks/Cole Publ. Co.},\ \bibinfo {year}
  {1998})\BibitemShut {NoStop}%
\bibitem [{\citenamefont {Site}(2019)}]{ICGEMWeb}%
  \BibitemOpen
  \bibfield  {author} {\bibinfo {author} {\bibfnamefont {ICGEM~Web}\
  \bibnamefont {Site}},\ }\href {http://icgem.gfz-potsdam.de/vis3d/longtime}
  {\enquote {\bibinfo {title} {{Visualization of Gravity Field Models and their
  Differences}},}\ } (\bibinfo {year} {2019})\BibitemShut {NoStop}%
\bibitem [{\citenamefont {White}(1986)}]{white1986world}%
  \BibitemOpen
  \bibfield  {author} {\bibinfo {author} {\bibfnamefont {Haschal~L}\
  \bibnamefont {White}},\ }\href
  {https://apps.dtic.mil/sti/citations/ADA166519} {\emph {\bibinfo {title} {The
  World Geodetic System 1984 - Earth Gravitational Model.}}},\ \bibinfo {type}
  {Tech. Rep.}\ (\bibinfo  {institution} {Defense Mapping Agency Aerospace
  Center St Louis Afs Mo},\ \bibinfo {year} {1986})\BibitemShut {NoStop}%
\bibitem [{\citenamefont {Hofmann-Wellenhof}\ and\ \citenamefont
  {Moritz}(2006)}]{hofmann2006physical}%
  \BibitemOpen
  \bibfield  {author} {\bibinfo {author} {\bibfnamefont {Bernhard}\
  \bibnamefont {Hofmann-Wellenhof}}\ and\ \bibinfo {author} {\bibfnamefont
  {Helmut}\ \bibnamefont {Moritz}},\ }\enquote {\bibinfo {title} {Physical
  geodesy},}\ \ (\bibinfo  {publisher} {Springer Science \& Business Media},\
  \bibinfo {year} {2006})\BibitemShut {NoStop}%
\bibitem [{197(1978)}]{1978135}%
  \BibitemOpen
  \bibfield  {title} {\enquote {\bibinfo {title} {{Chapter 6 Gravity
  Reductions, Corrections, and Anomalies}},}\ }in\ \href {\doibase
  https://doi.org/10.1016/S0422-9894(08)71156-5} {\emph {\bibinfo {booktitle}
  {Marine Gravity}}},\ \bibinfo {series} {Elsevier Oceanogr. Ser.},
  Vol.~\bibinfo {volume} {22},\ \bibinfo {editor} {edited by\ \bibinfo {editor}
  {\bibfnamefont {Peter}\ \bibnamefont {Dehlinger}}}\ (\bibinfo  {publisher}
  {Elsevier},\ \bibinfo {year} {1978})\ pp.\ \bibinfo {pages}
  {135--164}\BibitemShut {NoStop}%
\bibitem [{\citenamefont {Ince}\ \emph {et~al.}(2019)\citenamefont {Ince},
  \citenamefont {Barthelmes}, \citenamefont {Rei{\ss}land}, \citenamefont
  {Elger}, \citenamefont {F{\"o}rste}, \citenamefont {Flechtner},\ and\
  \citenamefont {Schuh}}]{ince2019icgem}%
  \BibitemOpen
  \bibfield  {author} {\bibinfo {author} {\bibfnamefont {E~Sinem}\ \bibnamefont
  {Ince}}, \bibinfo {author} {\bibfnamefont {Franz}\ \bibnamefont
  {Barthelmes}}, \bibinfo {author} {\bibfnamefont {Sven}\ \bibnamefont
  {Rei{\ss}land}}, \bibinfo {author} {\bibfnamefont {Kirsten}\ \bibnamefont
  {Elger}}, \bibinfo {author} {\bibfnamefont {Christoph}\ \bibnamefont
  {F{\"o}rste}}, \bibinfo {author} {\bibfnamefont {Frank}\ \bibnamefont
  {Flechtner}}, \ and\ \bibinfo {author} {\bibfnamefont {Harald}\ \bibnamefont
  {Schuh}},\ }\bibfield  {title} {\enquote {\bibinfo {title} {{ICGEM--15 years
  of successful collection and distribution of global gravitational models,
  associated services, and future plans}},}\ }\href {\doibase
  https://doi.org/10.5194/essd-11-647-2019} {\bibfield  {journal} {\bibinfo
  {journal} {Earth System Science Data}\ }\textbf {\bibinfo {volume} {11}},\
  \bibinfo {pages} {647--674} (\bibinfo {year} {2019})}\BibitemShut {NoStop}%
\bibitem [{\citenamefont {Torres-Payoma}\ \emph {et~al.}(2020)\citenamefont
  {Torres-Payoma}, \citenamefont {Escobar}, \citenamefont {Castro},
  \citenamefont {Triana},\ and\ \citenamefont {Herrera}}]{ceur}%
  \BibitemOpen
  \bibfield  {author} {\bibinfo {author} {\bibfnamefont {Freddy}\ \bibnamefont
  {Torres-Payoma}}, \bibinfo {author} {\bibfnamefont {Manuel}\ \bibnamefont
  {Escobar}}, \bibinfo {author} {\bibfnamefont {Leyton}\ \bibnamefont
  {Castro}}, \bibinfo {author} {\bibfnamefont {Karla}\ \bibnamefont {Triana}},
  \ and\ \bibinfo {author} {\bibfnamefont {Diana}\ \bibnamefont {Herrera}},\
  }\bibfield  {title} {\enquote {\bibinfo {title} {Use and design of virtual
  and remote free access experiments: World pendulum alliance and dlab in times
  of covid 19},}\ }\href {http://ceur-ws.org/Vol-3037/paper6.pdf} {\bibfield
  {journal} {\bibinfo  {journal} {CEUR-WS Proceddings}\ } (\bibinfo {year}
  {2020})}\BibitemShut {NoStop}%
\bibitem [{\citenamefont {Instituto
  Superior~Técnico}(2012{\natexlab{a}})}]{elab}%
  \BibitemOpen
  \bibfield  {author} {\bibinfo {author} {\bibfnamefont {University of~Lisboa}\
  \bibnamefont {Instituto Superior~Técnico}},\ }\href
  {https://www.e-lab.ist.utl.pt/} {\enquote {\bibinfo {title} {{e-Lab remote
  laboratories}},}\ } (\bibinfo {year} {2012}{\natexlab{a}})\BibitemShut
  {NoStop}%
\bibitem [{\citenamefont {ORACLE}(2022)}]{java}%
  \BibitemOpen
  \bibfield  {author} {\bibinfo {author} {\bibnamefont {ORACLE}},\ }\href
  {https://www.java.com/es/download/} {\enquote {\bibinfo {title} {{Download
  Java}},}\ } (\bibinfo {year} {2022})\BibitemShut {NoStop}%
\bibitem [{\citenamefont {organization}(2022)}]{vlc}%
  \BibitemOpen
  \bibfield  {author} {\bibinfo {author} {\bibfnamefont {VideoLAN}\
  \bibnamefont {organization}},\ }\href
  {https://www.videolan.org/vlc/index.es.html} {\enquote {\bibinfo {title}
  {{VLC media player: Download}},}\ } (\bibinfo {year} {2022})\BibitemShut
  {NoStop}%
\bibitem [{\citenamefont {Instituto
  Superior~Técnico}(2012{\natexlab{b}})}]{elabweb}%
  \BibitemOpen
  \bibfield  {author} {\bibinfo {author} {\bibfnamefont {University of~Lisboa}\
  \bibnamefont {Instituto Superior~Técnico}},\ }\href
  {http://elab.ist.utl.pt/rec.web//} {\enquote {\bibinfo {title} {{laboratories
  e-lab: Download}},}\ } (\bibinfo {year} {2012}{\natexlab{b}})\BibitemShut
  {NoStop}%
\bibitem [{\citenamefont {Taylor}(1997)}]{taylor1997introduction}%
  \BibitemOpen
  \bibfield  {author} {\bibinfo {author} {\bibfnamefont {John}\ \bibnamefont
  {Taylor}},\ }\enquote {\bibinfo {title} {{Introduction to error analysis: The
  study of uncertainties in physical measurements}},}\ \ (\bibinfo  {publisher}
  {Univ. Science Books},\ \bibinfo {year} {1997})\BibitemShut {NoStop}%
\end{thebibliography}%
\end{document}